\documentclass[a4paper,11pt]{article}

\usepackage{booktabs} % for toprule, midrule, bottomrule in tables
\usepackage{authblk} % for multiple authors and affiliations
\usepackage{amsmath}
\usepackage{hyperref}
\usepackage{cleveref}
\usepackage{autonum}
\usepackage{amsfonts,amssymb,mathtools}
\usepackage{xcolor}

\usepackage[top=3cm,left=2cm,right=2cm,bottom=2cm]{geometry}

\providecommand{\keywords}[1]
{
  \small	
  \textbf{\textit{Keywords:}} #1
}

\usepackage{graphicx,float}
\usepackage{multirow}
\usepackage{bm}
\usepackage[font={small,it}]{caption}	
\usepackage[authoryear]{natbib}
\DeclareMathOperator*{\argmax}{arg\,max}

\date{}

\author[1]{Thais Pacheco Menezes}
\author[1,2,3,4,*]{Thomas Brendan Murphy}
\author[1,*]{Michael Fop}
\affil[1]{School of Mathematics and Statistics, University College Dublin}
\affil[2]{Insight Centre for Data Analytics, University College Dublin}
\affil[3]{Institut d'\'{E}tudes Avanc\'{e}es, Universit\'{e} de Lyon}
\affil[4]{ERIC, Universit\'{e} de Lyon}
\affil[*]{Thomas Brendan Murphy and Michael Fop have contributed equally to this work}
\affil[ ]{Corresponding author:thais.pachecomenezes@ucdconnect.ie}

\title{Hausdorff Distance-Based Record Linkage for Improved Matching of Households and Individuals in Different Databases}

\begin{document}

\maketitle

\begin{abstract}
Matching households and individuals across different databases poses challenges due to the lack of unique identifiers, typographical errors, and changes in attributes over time. Record linkage tools play a crucial role in overcoming these difficulties. This paper presents a multi-step record linkage procedure that incorporates household information to enhance the entity-matching process across multiple databases. Our approach utilizes the Hausdorff distance to estimate the probability of a match between households in multiple files. Subsequently, probabilities of matching individuals within these households are computed using a logistic regression model based on attribute-level distances. These estimated probabilities are then employed in a linear programming optimization framework to infer one-to-one matches between individuals. To assess the efficacy of our method, we apply it to link data from the Italian Survey of Household Income and Wealth across different years. Through internal and external validation procedures, the proposed method is shown to provide a significant enhancement in the quality of the individual matching process, thanks to the incorporation of household information. A comparison with a standard record linkage approach based on direct matching of individuals, which neglects household information, underscores the advantages of accounting for such information. 


\end{abstract}

\section{Introduction}

Record linkage is the process of matching information from different sources that are believed to be related to the same entity \citep{herzog2007data}. Due to the digitization of census and survey-based data collection approaches, the application of record linkage methods to match entries across different databases is a field of growing interest, which enables the investigation of changes in population, demographic patterns, and family transitions over time \citep{ruggles:2018,abramitzky:2020,abramitzky:2021,helgertz:2022}. The challenges regarding this task are related to the fact that the matching procedure is often based on information reported by the entity in the study, a process highly subject to typographical errors, inconsistencies, and changes over time \citep{abramitzky:2021}. Furthermore, errors in the data may appear as a consequence of how the survey was designed \citep{MesurementErrorSHIW}.

Concerning general record linkage methodology, a large body of work has been produced. One of the most famous approaches is the Fellegi-Sunter model \citep{fellegi1969theory}, which is widely used \citep[e.g.][]{sadinle2013generalized}, and has also been recently implemented in conjunction with a linear programming framework \citep{moretti2019optimization}. Bayesian extensions of the Fellegi-Sunter model and other approaches for record linkage have also been proposed in recent years \citep[e.g.][] {steorts2016bayesian, sadinle2017bayesian, tancredi:liseo:2011, fortini2001bayesian}. Other approaches consider graph matching methods: \citet{papadakis2022bipartite}, for example, compares the performance of multiple bipartite graph matching algorithms.

One of the motivations behind our novel contribution derives from the fact that most of the widely used record linkage methods are focused solely on matching individuals, ignoring any available grouping information, such as household membership. The incorporation of such information in the linkage process has not been extensively explored. \citet{frisoli2018exploring} investigate the effect of including household information when matching records, comparing the results of record linkage procedures with and without such information. The model proposed by \citet{fu2014graph} uses a graph matching framework which is shown to improve linkage accuracy when including the complete household structure in the matching process. Another work that considers household information is the one of \citet{fu2011automatic}, where the main idea is to use the household membership to clean and link the data in a way that records containing errors and variations can be corrected, reducing the number of wrongly matched entities. Record linkage with grouping information is also explored by \citet{on2007group}, who propose a metric to measure similarity between groups that allows eliminating sets unlikely to be matching, enabling to focus the matching of entities in those groups having high similarity. Following this theme of including household information when matching entities, we define a general multi-step record linkage procedure that allows the incorporation of household information to improve the process of matching records across different databases. The methodology is developed and illustrated in application to record linkage of the Bank of Italy Survey of Household Income and Wealth (SHIW) databases \citep{URL}.

An important step in matching databases is the quantification of the dissimilarity or similarity between pairs of records. Different metrics are employed to measure the dissimilarity between records, which are normally computed on the variables the entities are matched upon. According to the nature of these variables (text, numerical, categorical, etc.), different metrics can be used \citep{cohen2003comparison,herzog2007data,sayers2016probabilistic}. For example, for string variables, the Jaro-Winkler metric \citep{winkler1990string} is the most commonly used. If the variables to be compared are categories chosen by the respondent, the comparison can be directly done so that the dissimilarity would be zero if the entities belong to the same category or one otherwise. When entities are grouped according to some structure, the standard distance metrics employed for record linkage need to be modified, as in this situation it is required to measure the dissimilarity between sets of individuals \citep[see][for a comprehensive overview of distance measures between two sets of points]{eiter1997distance}. In our proposed methodology, the membership to a given household is incorporated in the matching process, and to quantify the dissimilarity of two households across databases, we propose the use of the Hausdorff distance \citep{hausdorffbook}. The Hausdorff distances between households are computed on the individual-level reported information and employed in a model used to predict the probability of a match between households. Subsequently, the matching of individuals is implemented by leveraging the information about the matched households and using a supervised learning method in combination with a linear programming optimization procedure. The use of the Hausdorff distance to incorporate household information is shown to be beneficial to the quality of the record linkage process. 

The paper is organized as follows: Section~\ref{s:data} describes the motivating Bank of Italy SHIW data; Section~\ref{s:methodology} presents the Hausdorff distance and the approach used to measure the dissimilarity between records, also introducing the supervised learning models employed to estimate matching probabilities between households and individuals within households; Section~\ref{s:household} presents and discusses the results of the process of matching households and individuals for the SHIW databases; Section~\ref{s:conclusion} concludes the paper with a discussion about limitations and potential developments.

\section{Data: Italian Survey of Household Income and Wealth} \label{s:data}

The Bank of Italy has been conducting the Italian Survey of Household Income and Wealth (SHIW) since 1960 to collect information about the incomes and savings of Italians. Over the years, the survey has been extended to include information about wealth, financial behaviour, and other general economic aspects. The study uses a sample drawn in two stages, with municipalities and households as the primary and secondary sampling units, respectively. The sample represents the population officially resident in Italy, not accounting for people living in institutions (convents, hospitals, prisons, etc.) or those who are in the country illegally. The size of the sample has gradually increased reaching about $8000$ households. In order to guarantee comparability across years, since $1989$, part of the sample comprises households that were interviewed in the previous study. In this way, about $50\%$ of the households are present in consecutive surveys. 

Data from the surveys are published approximately every two years on the Bank of Italy website \citep{URL}. The published microdata do not contain any information that could lead indirectly to the identification of the respondent, and, at the moment, they are available for download since $1989$ in different formats. Additional material, containing the questionnaire and data description, is also provided.

Due to the nature of the survey procedure, the true match status between households is available via the unique identifier assigned to each household the first time it was included in the study and retained for all future inclusions. For individuals, their matching status can be validated due to the presence of individual IDs in prior surveys. These will enable the assessment of the matching performance of the proposed method in a supervised manner. An important point to highlight is that only matches between individuals who remained in the same household can be detected since an individual's ID is solely associated with their household. This constraint emerges from the data collection procedure, as transitioning to a new household, like through marriage, necessitates the creation of a new individual ID, thus erasing any previous linkage. 

The record linkage framework proposed here is shown in application to the SHIW databases of $2014$, $2016$, and $2020$. The $2014$ database includes $19366$ individuals spread across $8156$ households, whereas the $2016$ database consists of $16462$ individuals within $7420$ households. The $2020$ database, on the other hand, comprises $15198$ individuals distributed among $6239$ households.

\begin{table}[b!]
\centering
\footnotesize
\caption{Description of the variables considered for the matching procedure available in the $2014$, $2016$, and $2020$ Italian survey.}
\label{t:variablesItaly}
\begin{tabular}{lll}
\hline
Variable & Description     & Range            \\
\hline \noalign{\smallskip}
SEX      & Individual's sex    & 2 levels      \\
CIT  & Indicator if an individual is an Italian citizen or not & 2 levels \\
ANASC    & Year of birth      & Discrete       \\
STUDIO   & Educational qualification & 8 levels \\
NASCREG  & Region of birth      & 21 levels     \\
NACE & Sector of activity of the company where the individual works/worked & 22 levels\\
IREG     & Region of residence  & 20 levels     \\
QUAL     & Employment status   & 7 levels    \\  \noalign{\smallskip}
\hline
\end{tabular}
\end{table}

The questionnaire used in the survey includes several variables. Table~\ref{t:variablesItaly} presents the name, the description, and the range of the variables considered in this work. Attention was given to variables conducive to constructing individual profiles, such as sex, year of birth, region of residence, and employment status. These variables offer a comprehensive view of the respondent's profile, although certain aspects may change over time, posing challenges in the matching process. Financial variables were omitted due to their sensitivity and volatility. In general, the majority of these variables are categorical, typically consisting of fewer than $10$ distinct levels. There are, however, a few exceptions: variables such as those indicating the region of birth (NASCREG) and residence (IREG), as well as the variable detailing the sector of activity of the individual's employer (NACE), exhibit a more extensive range with over $20$ levels. Additionally, the variable representing the year of birth (ANASC) is the only numerical variable in the data, characterized by a discrete range.

The databases include some missing values, generated under a not-missing-at-random mechanism. For the variable NASCREG, recording the region of birth, the absence of an answer is related to individuals not born in Italy. Hence, the missing entry is replaced by an extra category indicating that a subject is not born in Italy. The variable that indicates the sector of activity of the company (NACE) also has missing values, with most cases being individuals with working status corresponding to unemployed or pensioner. Also in this case a new category reflecting this information is created. However, eleven cases in $2016$ still include missing information. These cases are missing at random since their working status is specified but no information about their activity sector is available. In practice, this means that, when compared with the other individuals, these eleven cases will be considered to have maximum dissimilarity with other cases for the variable NACE.

\begin{figure}[t]
    \centering
 \includegraphics[width=12cm]{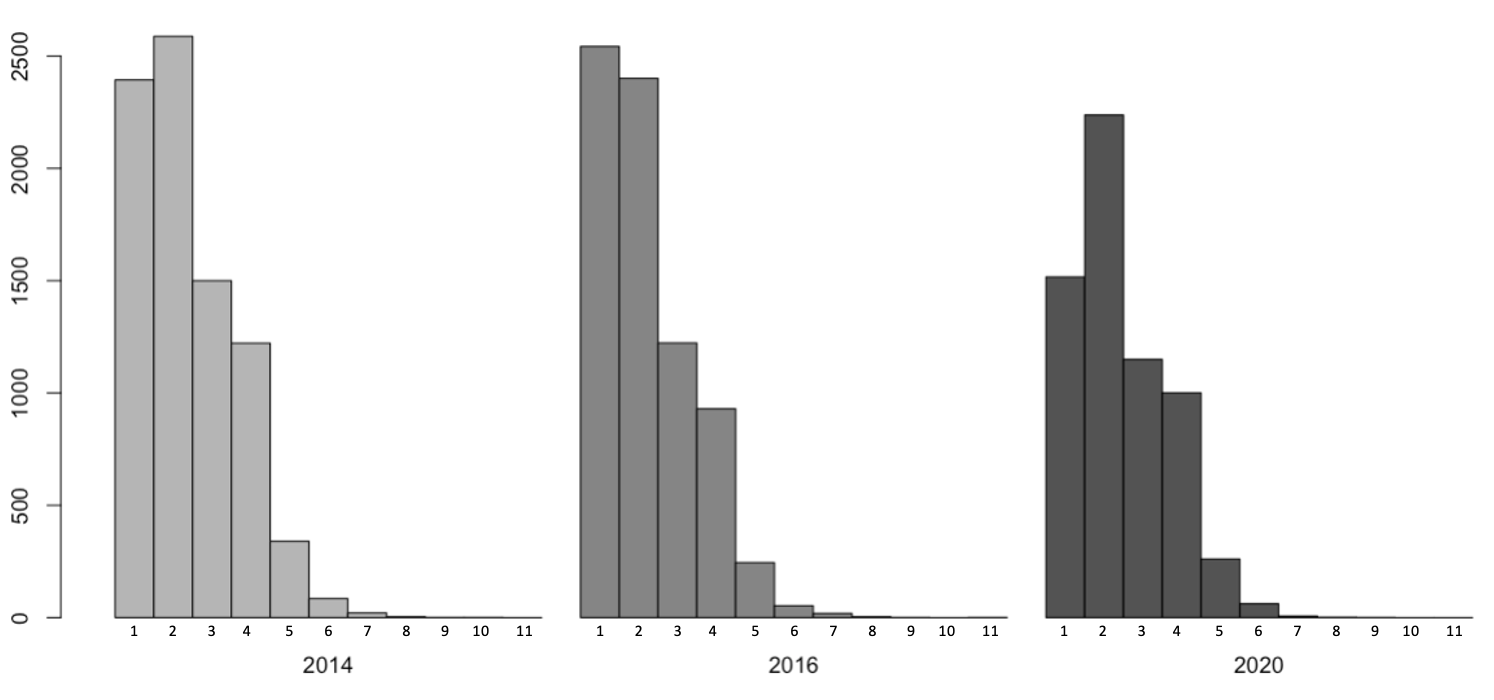}
\caption{Barplot for the distribution of the size of the households in the $2014$, $2016$, and $2020$ Italian survey.}
        \label{f:householdsize}
\end{figure}

Across the $2014$ and $2016$ datasets, the SHIW data include $3804$ matched households and $8660$ matched individuals. For the $2016$ and $2020$ datasets, a total of $2983$ household matches and $6434$ individual matches are present. In the data, for most matched households, the size of the households across survey years remains the same. Figure~\ref{f:householdsize} presents the barplot illustrating the distribution of household sizes for the years $2014$, $2016$, and $2020$. A comparative analysis reveals a reduction in the count of single-person households within the $2020$ survey. The growth in household size is evident in the calculated average household sizes: $2.37$ for $2014$, $2.22$ for $2016$, and an increase to $2.44$ for $2020$. However, it is worth noting that, aside from this shift, the overall distribution of household sizes remains relatively consistent across the three examined years with the majority of households being formed by up to two individuals. Regarding the area of residence, among the matched households across $2014$ and $2016$, only three changed their region of residence between surveys. This number increases to seven when comparing matched households between $2016$ and $2020$. These considerations underscore that the household structure tends to remain stable over survey years, and, given that most households are composed of two or more individuals, the inclusion of household information proves to be a valuable factor in the matching process.

It is essential to acknowledge that the SHIW data may be susceptible to errors and inconsistencies across survey years. A study by \citeauthor{MesurementErrorSHIW} (\citeyear{MesurementErrorSHIW}) delves into data quality and measurement errors in the SHIW data, revealing that inconsistencies can arise in responses due to factors like interview duration, the interview process, and the broader survey design. These inconsistencies can pose challenges in identifying matching entities accurately. Thus, incorporating multiple sources of information at both household and individual levels could prove advantageous for enhancing record linkage accuracy.

The subsequent section introduces the record linkage framework, which is exemplified using the SHIW databases of $2014$, $2016$, and $2020$. Our model primarily builds upon the $2014$ and $2016$ surveys, serving as foundational data for both the training and testing phases. Meanwhile, the $2020$ database is exclusively employed in a specific test scenario, as elucidated later. 

\section{Record Linkage with Household Information} \label{s:methodology}

We propose the {\em household-and-Hausdorff-distance-based record linkage method}, \texttt{hhlink}, a two-step record linkage framework that uses the Hausdorff distance as the input to a supervised learning method for matching households between databases. The matched households are then employed as the basis for matching individuals using a combination of supervised learning and linear programming optimization. In this section, we present the main components of \texttt{hhlink}. Since the training process employs the $2014$ and $2016$ databases, we will illustrate the methodology using these survey years.

\subsection{Hausdorff Distance Between Households} 

The Hausdorff distance \citep{hausdorffbook} measures the distance between two sets of points and it states that two sets are close if every point of either set is close to some point of the other one \citep{eiter1997distance}. 

Consider that the individual $i$ comes from the household ${\cal H}_s$ from the set of all households in the $2014$ database ${\cal H}^{2014}$, while the individual $j$ is from household ${\cal H}_t$ from the set ${\cal H}^{2016}$ of all the households in the $2016$ survey file. The distance between individuals $i$ and $j$ for a feature $k$ is denoted with $d_{ijk}$. Consider $\beta_1, \ldots, \beta_k, \ldots, \beta_K$ a collection of non-negative coefficients. The linear combination of all $K$ feature distances between individuals $i$ and $j$ is defined as

\begin{equation} \label{eq:linearcombination}
    d_{ij} = \sum_{k=1}^{K} \beta_k d_{ijk}.
\end{equation} 
The Hausdorff distance between households ${\cal H}_s$ and ${\cal H}_t$ is then defined as: 

\begin{equation} \label{eq:Hausdorff}
        \Delta_{s t} = \max\left\{ \max_{i \in {\cal H}_s}\min_{j \in {\cal H}_t} d_{ij}; \max_{j \in {\cal H}_t}\min_{i \in {\cal H}_s} d_{ij}\right\}.
\end{equation}

The Hausdorff distance corresponds to the greatest distance from any individual in one household to the most similar individual in the other household. Therefore, two households are considered close and likely to be a match if every individual in a household is close to the individuals in the other household. Consequently, pairs of households with the lowest Hausdorff distance are more likely to include matching individuals. Hence, the identification of similar households will enhance the process of linking individuals. The estimation of the $\beta_k$ coefficients is implemented via maximum likelihood, as it is explained in Section~\ref{s:householdmodel}.

For categorical variables, the distance $d_{ijk}$ for two individuals is $0$ if the individuals belong to the same category and $1$ otherwise. For the numeric variable year of birth, the distance is calculated as the absolute difference between the years as $d_{ijk} = {|{ANASC}_i - {ANASC}_j|}/{50}$, where ${ANASC}_i$ and ${ANASC}_j$ are the year of birth of individual $i$ and of individual $j$ respectively; the factor $50$ is to scale this distance so that it is close in range to $0$ and $1$. It is important to highlight that this factor is considered for interpretability only and does not affect the results. 

\subsection{Household Model} \label{s:householdmodel}

Let $y_{st}$ be the indicator binary variable of matching status between households, taking the value of $1$ if households ${\cal H}_s$ and ${\cal H}_t$ are a match and $0$ otherwise. We are interested in determining the probability $p_{st}$ of a match between the households  ${\cal H}_s$ and ${\cal H}_t$. The household model defines this probability as a logistic function dependent on the Hausdorff distance \eqref{eq:Hausdorff} between households:

\begin{equation} \label{eq:probability}
p_{st} = P(y_{st} = 1 | \Delta_{st}) = \frac{ e^{\beta_0 - \Delta_{st}} }{ 1 + e^{\beta_0 - \Delta_{st}}} , \qquad \text{with} ~ \beta_k \geq 0 ~ \text{ for $k$ } = 1, \dots, K,
\end{equation}
where $\beta_0$ is the intercept.  Note that the coefficients $\beta_k$ are in the linear combination in the term $\Delta_{st}$. Grouping all the parameters into the vector $\bm{\beta} = (\beta_0, \beta_1, \ldots, \beta_k, \ldots, \beta_K)$ and considering a Bernoulli model, the log-likelihood for two databases is given as follows:

\begin{equation} 
    \ell(\bm{\beta}) = \sum_{s = 1}^{N^{2014}} \sum_{t = 1}^{N^{2016}} \left\{{y_{st}} \log p_{st} + {(1-y_{st})}\log(1-p_{st})\right\},
\end{equation}
where $N^{2014}$ and $N^{2016}$ are the total number of households in $2014$ and $2016$ respectively.

The model is estimated and assessed within a supervised learning framework, hence consider a partition of the available data into a training and a test set. The procedure for dividing the data into training and test sets will be elucidated in Section~\ref{s:traintest}. During the training process, the true match status $y_{st}$ between households is known, and the parameter vector $\bm{\beta}$ is estimated through maximum likelihood:

\begin{equation}
    \bm{\hat{\beta}} = \argmax_{\bm{\beta}} \ell(\bm{\beta}) = \sum_{s = 1}^{N^{2014}} \sum_{t = 1}^{N^{2016}} \left\{{y_{st}} \log p_{st} + {(1-y_{st})}\log(1-p_{st})\right\}, \quad \text{with} ~ \beta_k \geq 0 ~ \text{ for $k$ } = 1, \dots, K.
\end{equation}
The above optimization is performed using standard box-constrained optimization via L-BFGS-B, as implemented in the R package \texttt{optimx} \citep{rsoftware, optimx1,optimx2, optimx}. Since we are dealing with distances, the model should encompass the fact that increasing the distance would cause the probability of a match to decrease. For this purpose, we impose the constraint that the $\beta_k$ parameters are non-negative ($\beta_k \geq 0$ for $k = 1, \dots, K$). It is important to highlight that this log-likelihood is a complex non-linear function of the $\bm{\beta}$ parameters throughout the Hausdorff distance $\Delta_{st}$ in \eqref{eq:Hausdorff} presented in the calculation of the probability of the household being a match \eqref{eq:probability}. 

Upon obtaining the estimated set of parameters $\bm{\hat{\beta}}$ we can proceed to calculate the Hausdorff distance between any two households in the test data, using the equation:

\begin{equation} \label{eq:HausdorffTest}
        \hat\Delta_{s t}^* = \max\left\{ \max_{i \in {\cal H}^*_s}\min_{j \in {\cal H}^*_t} d_{ij}^*; \max_{j \in {\cal H}^*_t}\min_{i \in {\cal H}^*_s}  d_{ij}^*\right\}.
\end{equation}
Here, $d_{ij}^* = \sum_{k=1}^{K} \hat\beta_k d_{ijk}^*$ represents the linear combination of the individual's distances in the test data for the $K$ features, taking into account the estimated parameter values $\hat{\beta}_k$. Following the computation of the Hausdorff distance, we can compute the estimated probability of a match between two households using the equation:

\begin{equation} \label{eq:probability2}
\hat{p}_{st}^* = P(\hat{y}_{st}^* = 1 | \hat\Delta_{s t}^*) = \frac{ e^{\hat\beta_0 - \hat\Delta_{s t}^*} }{ 1 + e^{\hat\beta_0 - \hat\Delta_{s t}^*}}.
\end{equation}
Utilizing these estimated probabilities, we can proceed to estimate the indicator $\hat{y}_{st}^*$, which denotes whether two households in the test data ${\cal H}_s^*$ and ${\cal H}_t^*$ are classified as a match. Specifically, $\hat{y}_{st}^*$ is assigned a value of $1$ if the estimated probability $\hat{p}_{st}^*$ associated with household ${\cal H}_t^*$ is the highest among all possible matches for household ${\cal H}_s^*$ and if $\hat{p}_{st}^* \geq \tau$. The threshold $\tau$ serves to control the proportion of matched households, enabling to determine that a household in one year does not have a match in the following survey if the highest probability of a match for that household is not sufficiently high. Spanning a range from $1$ to $0$, $\tau$ is defined as the highest value according to which the estimated proportion of matched households in the training data is as close as possible to the true proportion of matching households in the training set. It is essential to highlight that $\tau$ is estimated in the training phase and subsequently employed to determine the match status of households in the test data. Moreover, $\tau$ being equal to zero would imply that all households in $2014$ will be matched with a household in $2016$, while a value of one would signify that no households will be matched. In general, a higher value of $\tau$ corresponds to a lower proportion of matched households.

\subsection{Individual Model} \label{s:individualmodel}

To estimate the probability of a match between individuals in linked households, an individual-level logistic regression model is employed. The model is defined in terms of the linear combination of distances between the individuals to be matched:

\begin{equation}
D_{ij} = \alpha_1 d_{ij1} + \cdots + \alpha_k d_{ijk} + \cdots + \alpha_K d_{ijK},
\end{equation}
where $d_{ijk}$ are the individual level distances between subjects $i$ and $j$ on variable $k$. The $\alpha_k$ parameters are the regression coefficients, which weigh the impact of the distance on the probability of a match between two individuals. Similar to the household model, to ensure that when increasing the distance the probability of a match must decrease, the $\alpha_k$ parameters are constrained to be non-negative.

To train this model, only the households in the training data that have a match ($y_{st} = 1$) are considered and all individuals inside these households are paired together. We define a binary variable $z_{ij}$ which takes the value of $1$ if individuals $i$ and $j$ are a match. The probability of a match for a pair of individuals from the two matching households, denoted as $q_{ij}$, is expressed as follows:

\begin{equation} \label{eq:indprob}
q_{ij} = P(z_{ij} = 1 | D_{ij}, y_{st} = 1) = \dfrac{ e^{\alpha_0 - D_{ij}} }{ 1 + e^{\alpha_0 - D_{ij}} }, \qquad \text{with} ~ \alpha_k \geq 0 ~ \text{ for $k$ } = 1, \dots, K,
\end{equation}
where $\alpha_0$ is the intercept. Defining $\bm{\alpha} = (\alpha_0, \alpha_1, \ldots, \alpha_k, \ldots, \alpha_K)$ the vector of parameters, the model log-likelihood can be written as

\begin{equation} \label{eq:indlikelihood}
    \ell(\bm{\alpha}) = \sum_{s = 1}^{N^{2014}} \sum_{t = 1}^{N^{2016}} y_{st} \sum_{i = 1}^{n^{2014}} \sum_{j = 1}^{n^{2016}} \left\{{z_{ij}} \log q_{ij} + {(1-z_{ij})}\log(1-q_{ij})\right\} I_{i \in {\cal H}_s}I_{j \in {\cal H}_t},
\end{equation}
in which the indicator variables $I_{i \in {\cal H}_s}$ and $I_{j \in {\cal H}_t}$ take the value one if the individuals $i$ and $j$ belong to the households ${\cal H}_s$ or ${\cal H}_t$, respectively, and zero otherwise; $n^{2014}$ is the total number of individuals in the database for the year $2014$, and $n^{2016}$ is the total number of individuals in the database for the year $2016$. The model training involves considering instances of matching individuals within matching households. 

The model incorporates a non-negativity constraint on the coefficients. Furthermore, a ridge penalty is considered to correct for quasi-separation \citep{albert1984existence,heinze2006comparative} and to obtain non-divergent and interpretable coefficient estimates. Consequently, to estimate the $\bm{\alpha}$ parameters, we maximize the following penalized log-likelihood:

\begin{equation} \label{eq:indlikelihood}
\begin{aligned}
    \bm{\hat{\alpha}} = \argmax_{\bm{\alpha}} \ell(\bm{\alpha}) =  & \sum_{s = 1}^{N^{2014}} \sum_{t = 1}^{N^{2016}} y_{st} \sum_{i = 1}^{n^{2014}} \sum_{j = 1}^{n^{2016}} \left\{{z_{ij}} \log q_{ij} + {(1-z_{ij})}\log(1-q_{ij}) \right\} I_{i \in {\cal H}_s}I_{j \in {\cal H}_t} + \lambda \sum_{k = 1}^K \alpha_k^2 , \\
    & {\text{subject to } \alpha_k \geq 0 ~ \text{ for $k$ } = 1, \dots, K}.
\end{aligned}
\end{equation}
This penalized log-likelihood is equivalent to a penalized logistic regression model with a response variable corresponding to the matching status of individuals. Hence, the estimation is implemented using the efficient routines available in the R package {\tt glmnet} \citep{glmnet_paper, glmnet}, which allows the inclusion of the ridge penalty and the non-negativity constraints. Tuning of the penalty hyperparameter $\lambda$ is performed using the default cross-validation procedure in the package.

With the estimated parameter vector $\hat{\bm{\alpha}}$, and conditioning upon the fact that the households ${\cal H}_s^*$ and ${\cal H}_t^*$ in the test data are predicted to be a match, we can estimate the probability of a match between two individuals inside these households, using,

\begin{equation} \label{eq:indprobtest}
\hat{q}^*_{ij} = P(z_{ij}^* = 1 | \hat{D}_{ij}^*, \hat y_{st}^* = 1) = \dfrac{ e^{\hat\alpha_0 - \hat D_{ij}^*} }{ 1 + e^{\hat\alpha_0 - \hat D_{ij}^*} },
\end{equation}
where $\hat D_{ij}^* = \hat\alpha_1 d_{ij1}^* + \cdots + \hat\alpha_k d_{ijk}^* + \cdots + \hat \alpha_K d_{ijK}^*$. This probability quantifies the likelihood of a match between the specific pair of individuals $i$ and $j$ in the test data, based on the estimated model parameters. Subsequently, we use these estimated probabilities, $\hat{q}^*_{ij}$, within a linear programming framework to compute the estimate $\hat{z}^*_{ij}$, taking the value of $1$ if a match predicted between individuals $i$ and $j$, and $0$ otherwise. The details of the linear programming framework are presented in the following section.

\subsection{Linear Programming} \label{s:linearprog}

The proposed record linkage approach can be seen as a linear programming framework where a matrix of weights is assigned to each pair. In this context, the identification of the matches can be performed by maximizing the probability of a match under the constraint that each individual can be matched at most with one other individual. Following \cite{moretti2019optimization}, we propose a similar approach to match individuals within households: the probabilities estimated from the individual-level logistic regression are used in a linear programming optimization framework to enforce one-to-one matches between individuals in the matched households across the databases. Consider two matched households ${\cal H}_s$, ${\cal H}_t$ with $N_s$ and $N_t$ individuals to be matched, respectively, and for which the estimated $\hat{y}_{st} = 1$. Let $\hat{q}_{ij}$ denote the estimated probability of a match for the pair of individuals $i$ and $j$, and let $z_{ij}$ be the binary indicator of whether individual $i$ from household ${\cal H}_s$ is a match for individual $j$ from household ${\cal H}_t$ to be estimated. The matching problem can be expressed as the following linear programming optimization problem:

\begin{equation} 
\label{eq:optm}
\begin{split}
\hat{z}_{ij} = \argmax_{z_{ij}}\sum_{i=1}^{N_s}\sum_{j=1}^{N_t} \hat{q}_{ij} z_{ij} \\
\text{subject to} & \sum_{i = 1}^{N_s} z_{ij} \leq 1 \quad \quad j = 1, \ldots, N_t,  \\
& \sum_{j = 1}^{N_t} z_{ij} \leq 1 \quad \quad i = 1,\ldots, N_s, \\
& \sum_{i = 1}^{N_s} z_{ij}\hat{q}_{ij} \geq \bar{q} \quad \quad j = 1, \ldots, N_t, \\ 
& \sum_{j = 1}^{N_t} z_{ij} \hat{q}_{ij} \geq \bar{q} \quad \quad i = 1,\ldots, N_s,
\end{split}
\end{equation}
where $\bar{q}$ is the average estimated probability of a match between all the individuals in the matched households, $\bar{q} = \sum_{i,j} q_{ij} / (N_s N_t)$. In practice, the first two constraints indicate that each individual in $2014$ can be matched with only up to one other individual in $2016$, and vice versa. The last two constraints prevent matches for certain individuals, accounting for potential household changes such as individuals leaving or joining a household in the time between surveys. 
A pair of individuals within a matched household will not be matched if their probability of a match is below the average probability of a match for all pairs within that household. This criterion ensures that only individuals with a probability exceeding the average are assigned matches, while others remain unlinked.

We note that the same linear programming framework is implemented to estimate the matching status $z_{ij}^*$ of individuals in the test data, using accordingly the corresponding estimated probabilities $\hat{q}^*_{ij}$ and household matching labels $\hat{y}^*_{st}$.

\section{Assessment of the Record Linkage Procedure} \label{s:assessment}
The \texttt{hhlink} approach proposed in the previous section is a supervised learning framework. Therefore, its performance is evaluated by considering a training and test splitting procedure of the SHIW data. Different metrics are employed for evaluation, and the performance of \texttt{hhlink} is compared with that one of a Fellegi-Sunter model employed to link all the individuals directly.

\subsection{Training and Test Data} \label{s:traintest}
In the training process, the true match status between households and individuals is available and employed to estimate the parameters of the \texttt{hhlink} two-step approach. In the testing stage, the estimated models are used to assess the framework's performance in linking households and individuals. For this purpose, the available SHIW data are partitioned into training and test sets. Due to the structure of the data, the linkage goal, and the data collection process, two distinct methods are explored to split the data. These correspond to {\em internal} and {\em external} validation, respectively.

The first approach serves as an essential step in internally validating our methodology. Here, the training set is composed by selecting $60\%$ of matching households and $60\%$ of non-matching households for both $2014$ and $2016$ databases. Specifically, from the pool of households in $2014$ that has a match in $2016$, $60\%$ are randomly chosen along with their corresponding matches. Simultaneously, among the households in the $2014$ data that do not have a corresponding match in the $2016$ database, $60\%$ are randomly chosen to complete the training data for that year. This process is then repeated for the $2016$ data, where $60\%$ of the non-matched households from that year are similarly selected to create the corresponding training data. The remaining non-selected households will form the test set. The record linkage approach is subsequently deployed to predict match statuses on both the training and test data. To ensure that the results obtained are not merely the result of favorable training and test data partitions, the procedure is replicated ten times, and the results presented will reflect the model's average performance across all of these partitions.

The external validation approach involves training the model using the complete data from the $2014$ and $2016$ surveys and subsequently testing its performance by matching the $2016$ database with the $2020$ database. This method ensures the absence of data leakage, as none of the information from the $2014$ database or the real matching status is employed in the testing phase. Moreover, there is no overlap across databases for training and testing, given that the testing phase includes an entirely new database whose instances are not present at all in the training phase. This validation approach provides a distinct demarcation between the training and testing datasets. Furthermore, by conducting external validation, we can assess the model's performance when a new survey is released, with the objective of matching the fresh data to the latest available survey.

\subsection{Performance Measures} \label{s:qualitymeasures}

Given that information on actually matching households and individuals is available in the data, standard metrics can be used to assess the proposed record linkage framework's predictive performance.

The class distribution of matching/non-matching instances is considerably unbalanced due to the nature of the record linkage problem. For this reason, we consider the $F_1$ score to measure the performance of the models:

\begin{equation} \label{eq:f1socre}
    F_1 = \frac{2(\mbox{precision} \times \mbox{recall})}{\mbox{precision} + \mbox{recall}} = \frac{2\mbox{tp}}{2\mbox{tp} + \mbox{fp} + \mbox{fn}},
\end{equation}
where $\mbox{tp}$ represents true positives, $\mbox{fp}$ stands for false positives, and $\mbox{fn}$ denotes false negatives. Values of this score close to $1$ indicate good performance. In the score calculation, precision measures the proportion of correct matches among all matches made, while recall reflects the percentage of correct matches considering the known true match status.

We consider also the false positive (FPR) and false negative (FNR) rates, defined as: 

\begin{equation} \label{eq:fprfnr}
\begin{aligned}
    \mbox{FPR} = \frac{\mbox{fp}}{\mbox{fp} + \mbox{tn}} \\
    \mbox{FNR} = \frac{\mbox{fn}}{\mbox{fn} + \mbox{tp}}.
\end{aligned}
\end{equation}
The FPR measures the proportion of records that were wrongly assigned as matches while the FNR corresponds to the proportion of missed matching instances; lower values in both metrics are preferable. 

These metrics will be utilized to evaluate the quality of the record linkage process and to compare the performance of the proposed {\tt hhlink} with a model that matches individuals directly without taking into account household information.

\subsection{Method for Comparison} \label{s:comparison}

The primary objective of this paper is to demonstrate the enhancement achieved in the individual matching process through the preliminary matching of households. To ascertain the degree of improvement in match quality resulting from the inclusion of household information, we will conduct a comparative analysis against an alternative linking methodology. This evaluation aims to contrast our proposed approach with an alternative method that matches individuals directly.

The considered alternative method implements the Fellegi-Sunter model, estimated using the Expectation-Maximization algorithm as implemented in the \texttt{fastLink} R package \citep{enamorado2019using, fastLink}; in what follows, we denote with \texttt{fastLink} the Fellegi-Sunter model as implemented in the R package.

Given the computationally intensive nature of direct individual matching in \texttt{fastLink}, significant computational resources and time are essential prerequisites. This arises from the necessity to check the match possibilities for all potential pairs, a task that can become overwhelmingly large. For example, matching the $2014$ and $2016$ data would require evaluating an enormous number of pairs, totaling $19366 \times 16462$. To alleviate this computational burden, blocking is applied, a common practice in record linkage scenarios.

Blocking partitions records into exclusive groups, such that only records within the same block are eligible for matching. This approach dramatically reduces the number of pairs that require evaluation \citep{comparisonblocking}. We consider blocking on two factors: gender and region of birth, striving to achieve a balance between block size and the total number of blocks. Consequently, only individuals with the same gender and born in the same region are considered potential matches when applying the {\tt fastLink} method.

\section{Record Linkage of the SHIW Databases}
\label{s:household}

As outlined in Section~\ref{s:householdmodel}, we evaluate the model's performance by utilizing the estimated parameters to predict the match status between households ${\cal H}_t$ and ${\cal H}_s$ in training and test sets. Within each pair of households identified as a match, we assess the classification performance of the individual's model, as detailed in Section~\ref{s:individualmodel}. In this context, the parameters derived from the logistic regression and the linear programming optimization process enable us to estimate the indicator variable of whether individuals $i$ and $j$ constitute a match.

\subsection{Internal Validation} \label{s:internal}
    
Initially, we present the outcomes obtained when the databases of $2014$ and $2016$ were randomly divided into training and test data ten times. We note that due to the tenfold repetition of model fitting and testing, we are able to assess not only the variability of the parameter estimates but also the variability in the model's performance metrics. 

\begin{table}[t]
\centering
\caption{Average estimates of the household model's parameters. The StDev represents the standard deviation associated with each estimate.}
\label{t:comparingParam}
\begin{tabular}{lrrrrrrrrr}
\hline \noalign{\smallskip}
    & Intercept & SEX  & ANASC & CIT  & STUDIO & NACE & NASCREG & IREG & QUAL \\ \hline \noalign{\smallskip} 
    Estimate           & -0.27      & 2.82 & 13.74 & 0.00 & 1.63   & 1.41 & 3.33    & 7.98 & 0.01 \\ \noalign{\smallskip}
    StDev & 0.06      & 0.07 & 0.60  & 0.00 & 0.04   & 0.06 & 0.07    & 3.06 & 0.01 \\  \noalign{\smallskip} \hline
\end{tabular}
\end{table}

Table~\ref{t:comparingParam} presents the average estimates of the model coefficients $\beta_k$ alongside the standard deviation of the estimates considering all splits. The estimates show that the variable accounting for the distance between the years of birth (ANASC) is the one with the largest weight in the calculation of the Hausdorff distance. We highlight that the values for this variable are continuous while all the others are equal to $1$ if they are exactly the same for two individuals or $0$ otherwise. The variable with the second largest weight is the variable indicating the region of residence of the household (IREG) while the indicator of Italian citizenship (CIT) does not contribute to the distance in the model. 

Concerning variability, for the majority of variables, the estimates tend to vary closely around the mean value. The variable IREG exhibits the highest standard deviation, signifying greater variability in the estimations across model replications. Notably, despite the low standard deviation for the employment status variable (QUAL), the combination of a low average estimate suggests a substantial variability in this estimate, with a coefficient of variation of $100\%$.

\begin{figure}[b!]
  \centering
  \includegraphics[width=0.32\textwidth]{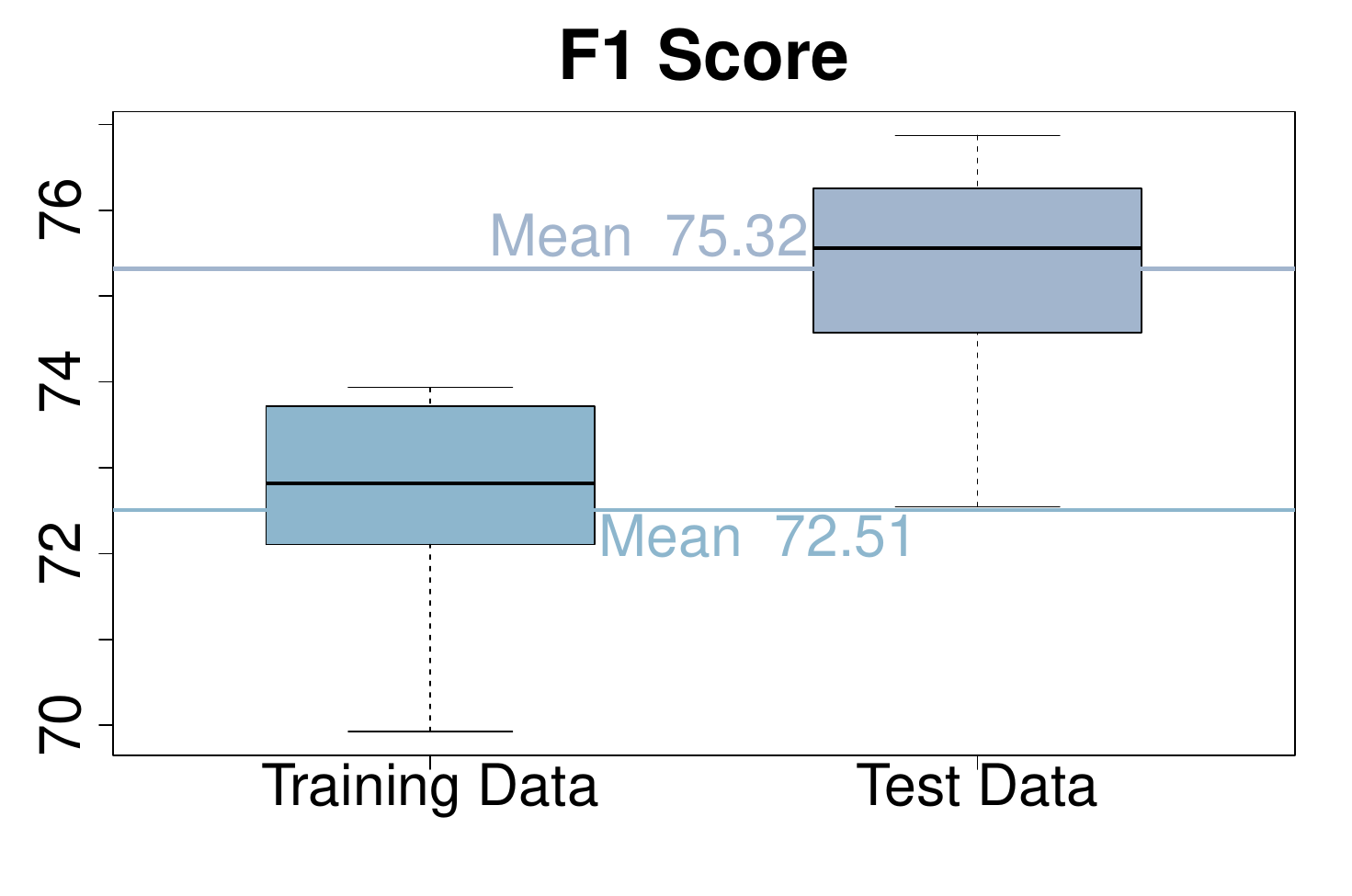}
  \centering
  \includegraphics[width=0.32\textwidth]{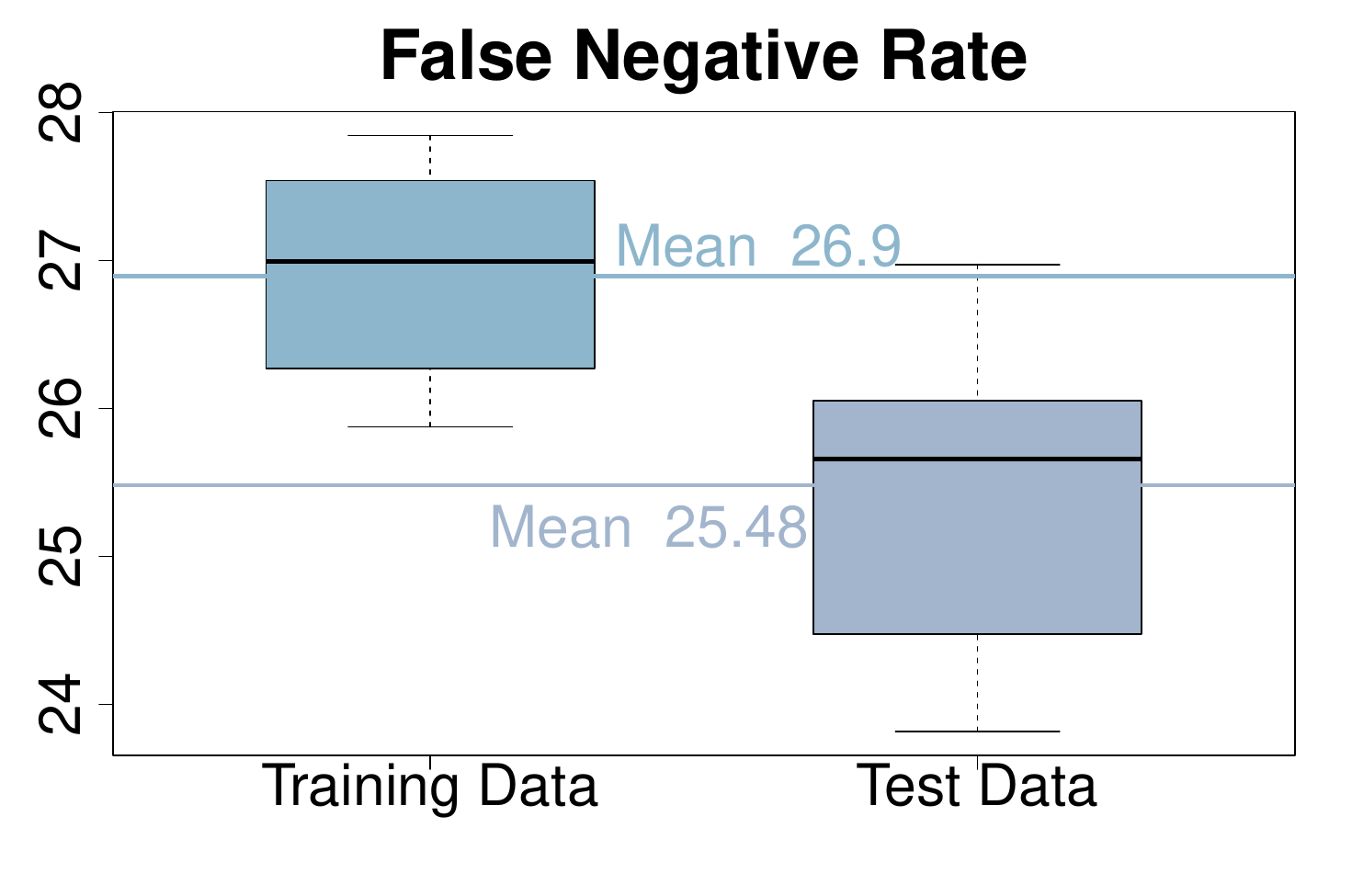}

  \centering
  \includegraphics[width=0.32\textwidth]{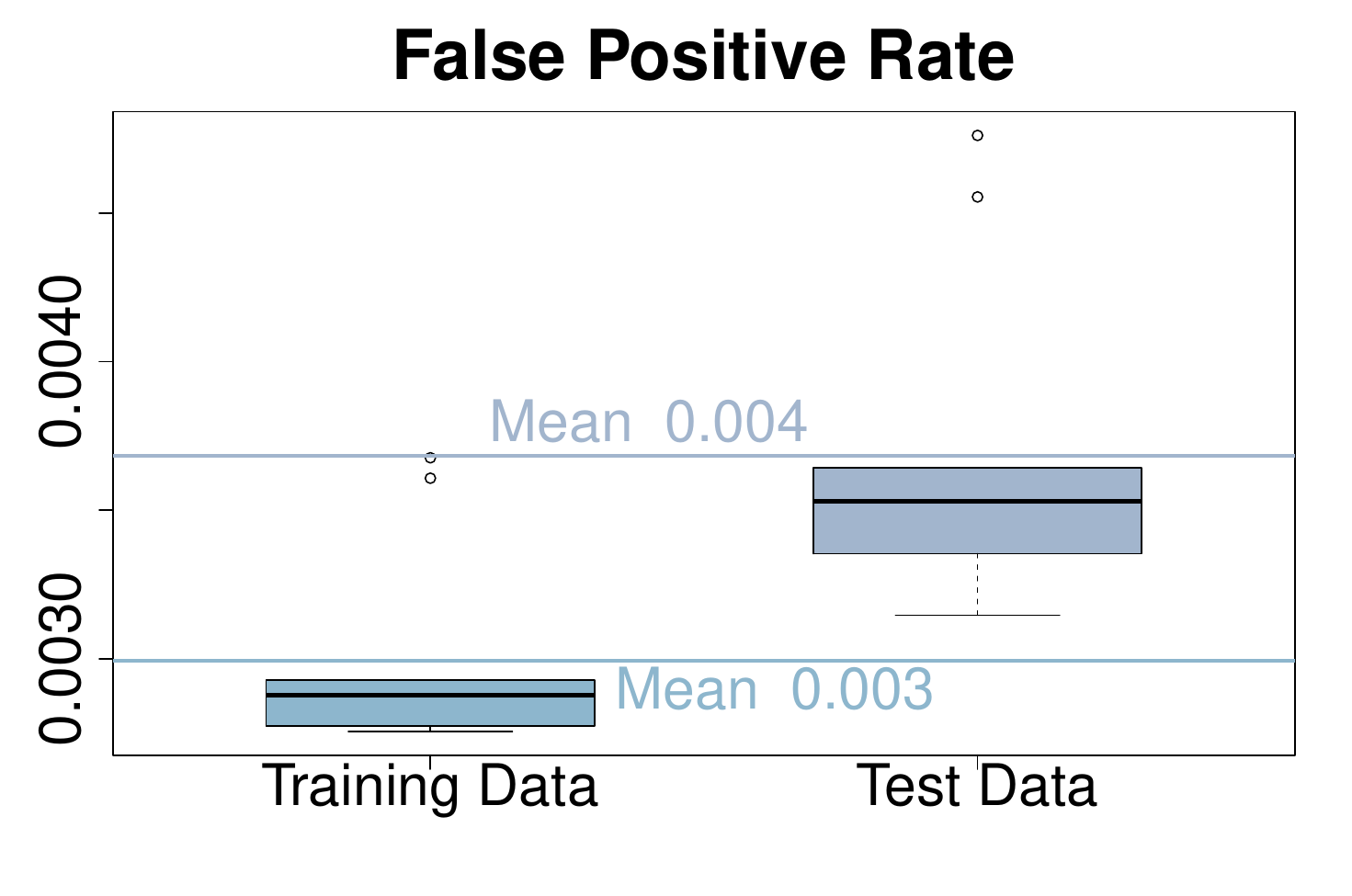}
  \centering
  \includegraphics[width=0.32\textwidth]{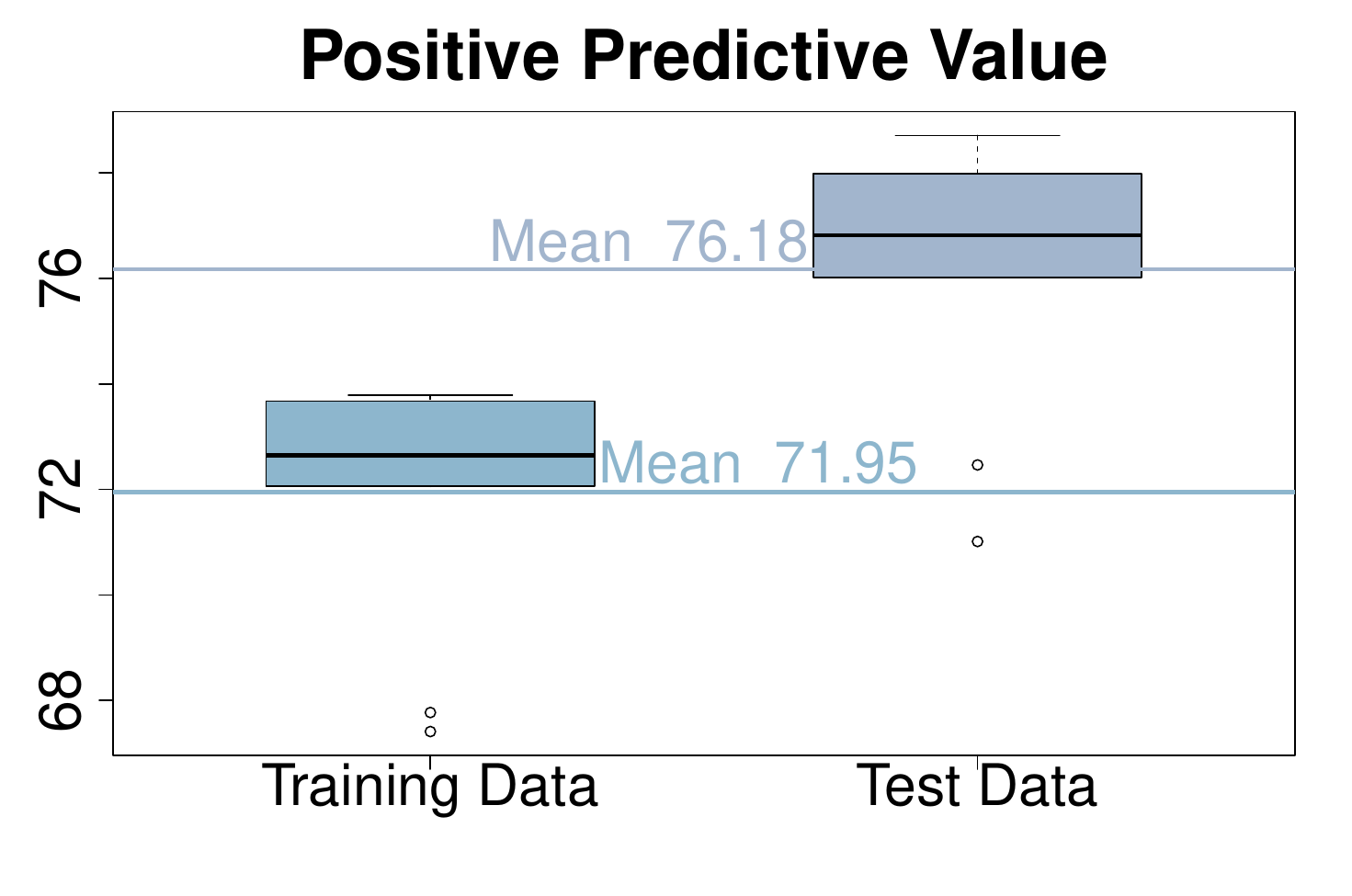}
  \centering
  \includegraphics[width=0.32\textwidth]{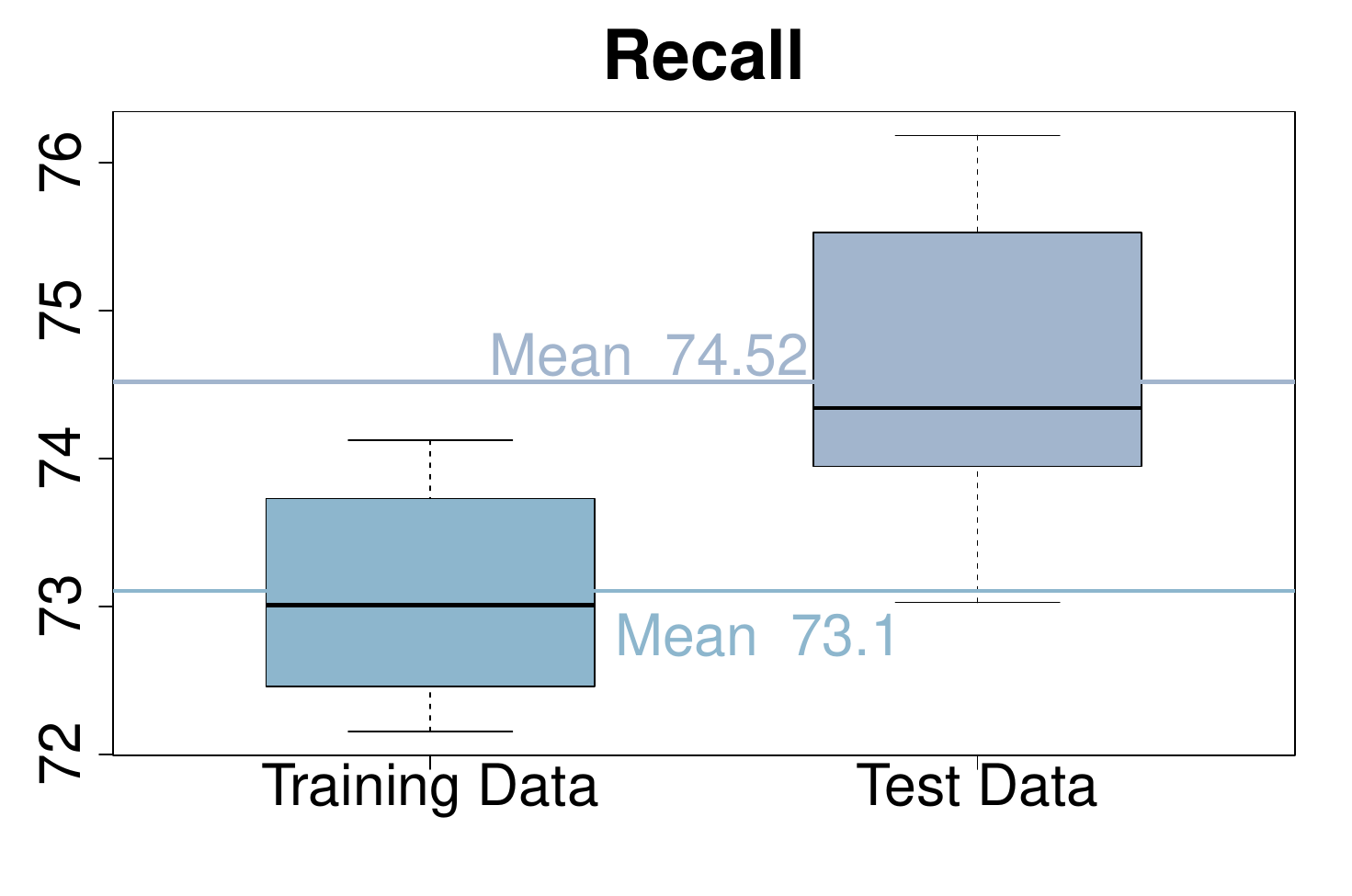}
\caption{Boxplots illustrating performance metrics for household pair match status prediction.
}
\label{f:household_metrics}
\end{figure}

Leveraging the available ground truth on matching households, Figure~\ref{f:household_metrics} presents quality measures for the household model considering all possible pairs of households. In this scenario, each household in $2014$ was paired with all households in the $2016$ set.
The results indicate a high $F_1$ Score, driven by high recall and positive predictive values, suggesting that, in general, the model can correctly classify household pairs. Specifically, the average recall of $73\%$ for the training data and $75\%$ for the test data underscores the model's ability to accurately identify household pairs known to be a match. It is important to note that the high false-negative rate may be attributed to situations where the highest probability of a match was not associated with the true matching household, or the estimated probability of a match fell below the defined threshold $\tau$. In the training data, $\tau$ is determined as the value that makes the estimated total number of matches between households approximately equal to the proportion of matches in the training data. Given an average matching proportion of $46.65\%$ in the training data, the estimated value for $\tau$ is $0.11$ on average. This implies that to achieve around $46\%$ matches between households in the training data, only households with probabilities greater than $0.11$ are considered as potential matches. The estimated $\tau$ value is then applied to the test data to filter the potential matches in the household model.

\begin{figure}[t!]
  \centering
  \includegraphics[width=0.45\textwidth]{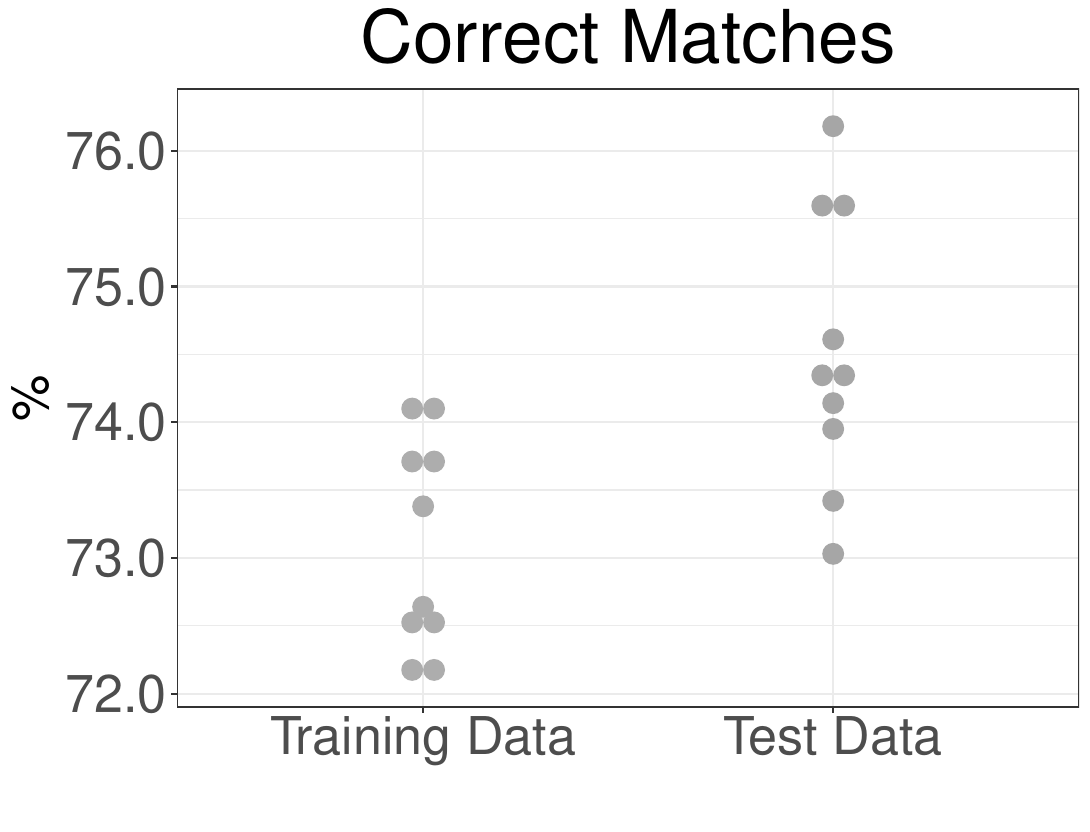}
  \centering
  \includegraphics[width=0.45\textwidth]{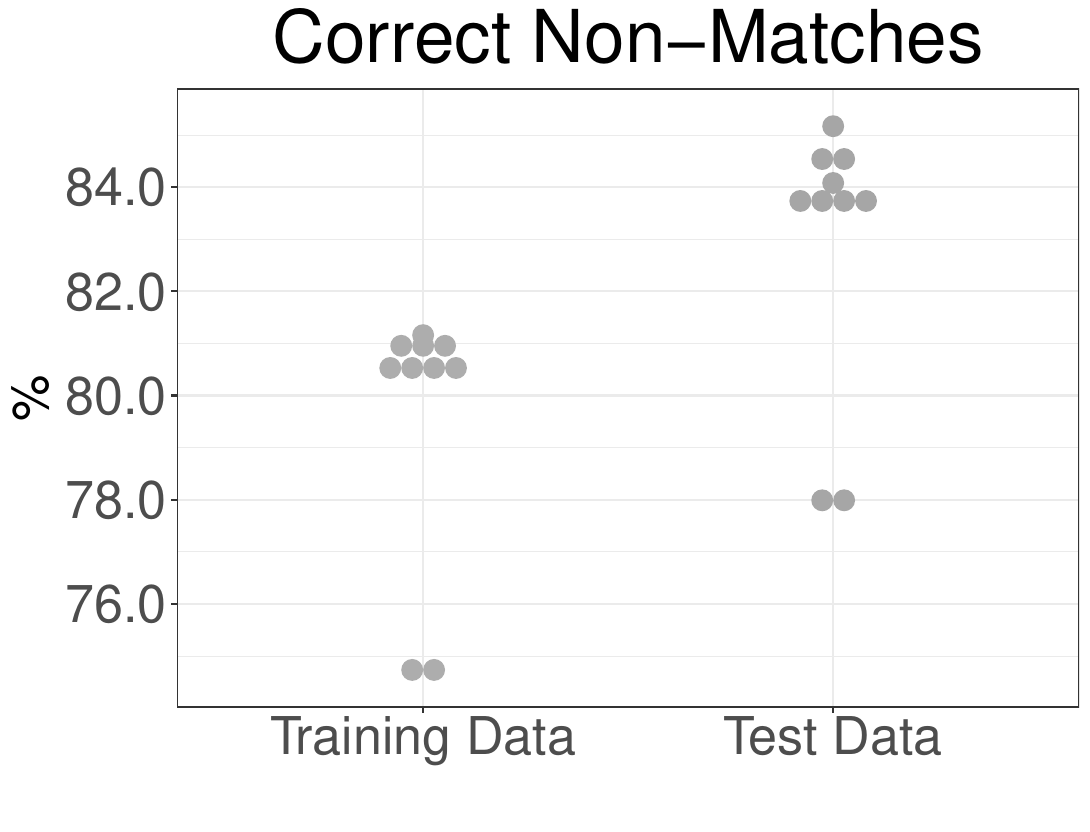}
\caption{The left plot illustrates the proportion of correctly matched households among those with a match in each data split. The right plot shows the proportion of households accurately identified as not having a match among those without a match.
}
\label{f:household_cm_cnm}
\end{figure}

We also assess the household model's performance by examining the predictive performance for each household in the $2014$ database individually, rather than considering all possible pairs. We examine whether the model correctly matches households that have a match and whether it accurately identifies households without a match for each household in $2014$. Figure~\ref{f:household_cm_cnm} illustrates, for each data partition, the proportion of correctly matched households (Correct Matches) and the proportion of households correctly classified as not having a match (Correct Non-Matches). The results demonstrate that, among households with a match, approximately $72\%$ to $74\%$ were correctly matched in the training set, expanding to $73\%$ to $76\%$ in the test data. For households without a match, the proportion of correct non-matches is higher, ranging from around $75\%$ to $81\%$ in the training data and $78\%$ to $85\%$ in the test data. These results focused on assessing the model's performance for each $2014$ household in the training and test data underscores the model's ability to effectively match these households or to accurately identify when they do not have a match.

To gain insights into the estimated probabilities of a match, we assess the rank of these probabilities for each household in $2014$ in relation to the corresponding true matches in the $2016$ database. For each split, the estimated probabilities of a match between a $2014$ household and all $2016$ households are ranked in descending order. Accurate model estimates would position the true matching household at the top. Table~\ref{t:positionmodel} presents the average percentage of actual matching households occupying the first position in the predicted probabilities, ranked in descending order, for both the training and test sets. The table shows that, on average, approximately $76.74\%$ of the highest probability from the model corresponds to the true match in the training data. For the test data, this value is $79.55\%$. Moreover, the top three positions correspond, on average, to actually matching pairs in around $85\%$ of the cases for both training and test sets. The results indicate that selecting the top three households with the highest probability of a match will likely include the true match in the selection.

\begin{table}[b!]
\centering
\caption{Rank of the probability of the correctly matching household. The rank $1$ indicates that the highest probability is associated with the $2016$ household which is the correct match. Likewise, rank $2$ implies that the match had the second-highest probability, and so on.}
\label{t:positionmodel}
\begin{tabular}{rrrrr}
\hline\noalign{\smallskip}
Rank & Training & Test \\ 
\noalign{\smallskip}
\hline
\noalign{\smallskip}
1        & 76.74 & 79.55 \\ 
2        & 4.92 & 4.19 \\ 
3        & 1.76  & 1.14  \\ 
4        & 1.09  & 0.64  \\ 
$\geq$ 5 & 15.49 & 14.48  \\ 
   \hline
\end{tabular}
\end{table} 

After having matched the households, the subsequent phase involves matching individuals within each linked household. Across each of the ten data splits into training and test subsets, conditionally on the matching households, we employ the logistic regression model outlined in Section~\ref{s:individualmodel} to predict the probability of a match between individuals. Table~\ref{t:individualparam} provides the average parameter estimates for the individual's model, along with their corresponding standard deviations. The variables associated with year of birth (ANASC) and sex (SEX) are the ones with the largest coefficients in the model, indicating that they contribute the most to the probability of two individuals being a match. The coefficient of the variable related to the region of residence (IREG) is shrunk to zero, in contrast to the household-level model where it had the second largest weight in the Hausdorff distance. This indicates that the region of residence largely contributes to linking two households, but it is no longer relevant when matching individuals within a pair of matched households. Regarding the variability of the estimates, the conclusion is similar to the one observed in the household model: most of the estimates do not vary too much around the mean.

\begin{table}[t!]
\centering
\caption{Average estimates of the individual's model parameters.}
\label{t:individualparam}
    \begin{tabular}{lrrrrrrrrr}
    \hline \noalign{\smallskip}
                       & Intercept & SEX  & ANASC & CIT  & STUDIO & NACE & NASCREG & IREG & QUAL \\ \hline \noalign{\smallskip} 
    Estimate           & 1.88      & 2.77 & 4.44 & 0.46 & 1.18   & 0.46 & 0.96    & 0.00 & 0.58 \\ \noalign{\smallskip}
    StDev & 0.02      & 0.02 & 0.02  & 0.10 & 0.02   & 0.02 & 0.03    & 0.00 & 0.02 \\  \noalign{\smallskip} \hline
    \end{tabular}
\end{table}

Given the estimated matched households and the fitted individual-level logistic regression model, the probabilities from the logistic regression are employed in an optimization framework, as explained in Section~\ref{s:linearprog}. When assessing the individual's model performance, we highlight that the results account only for the pair of individuals inside matched households.

\begin{figure}[b!]
\centering
  \includegraphics[width=0.32\textwidth]{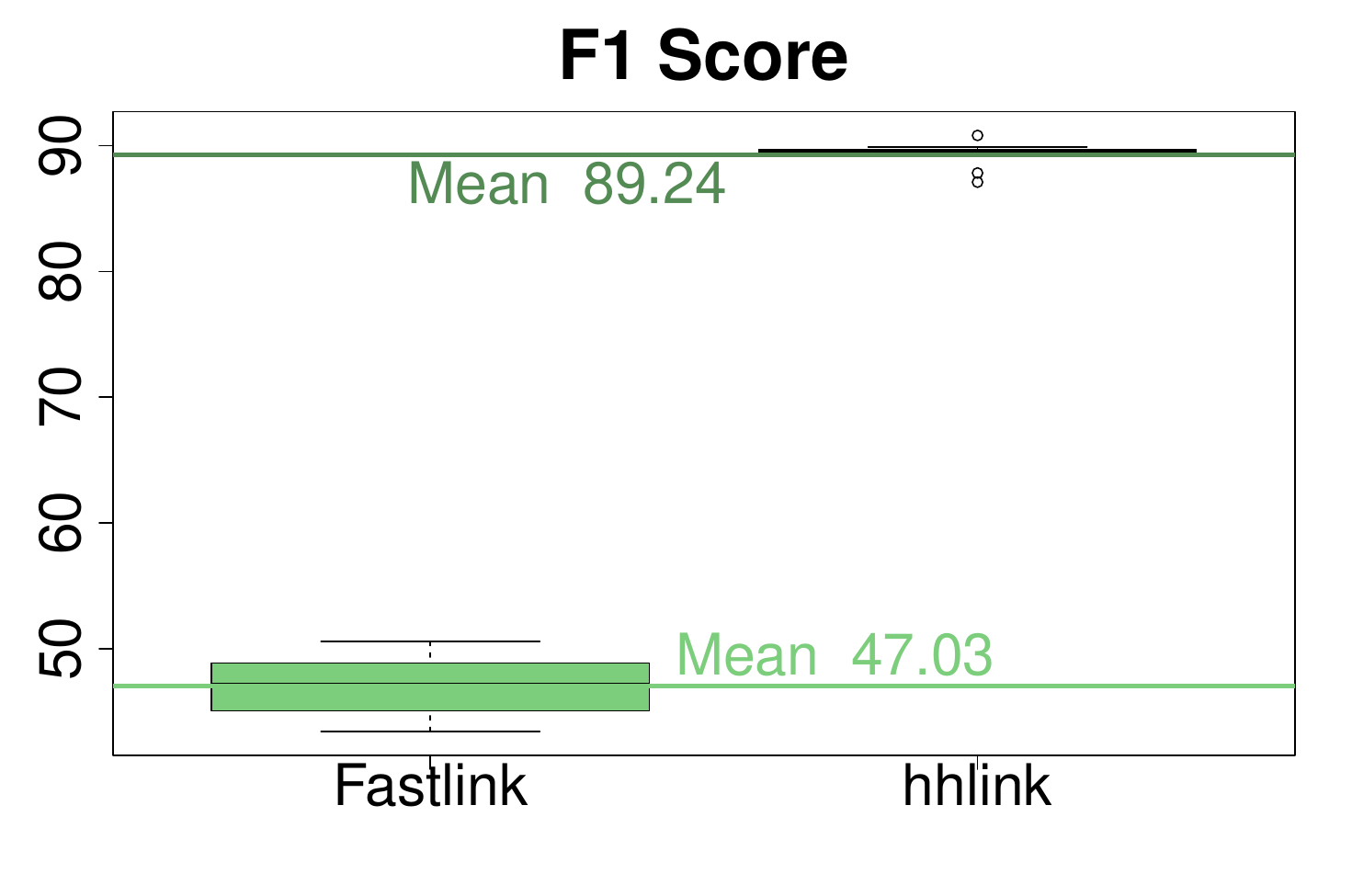}
  \centering
  \includegraphics[width=0.32\textwidth]{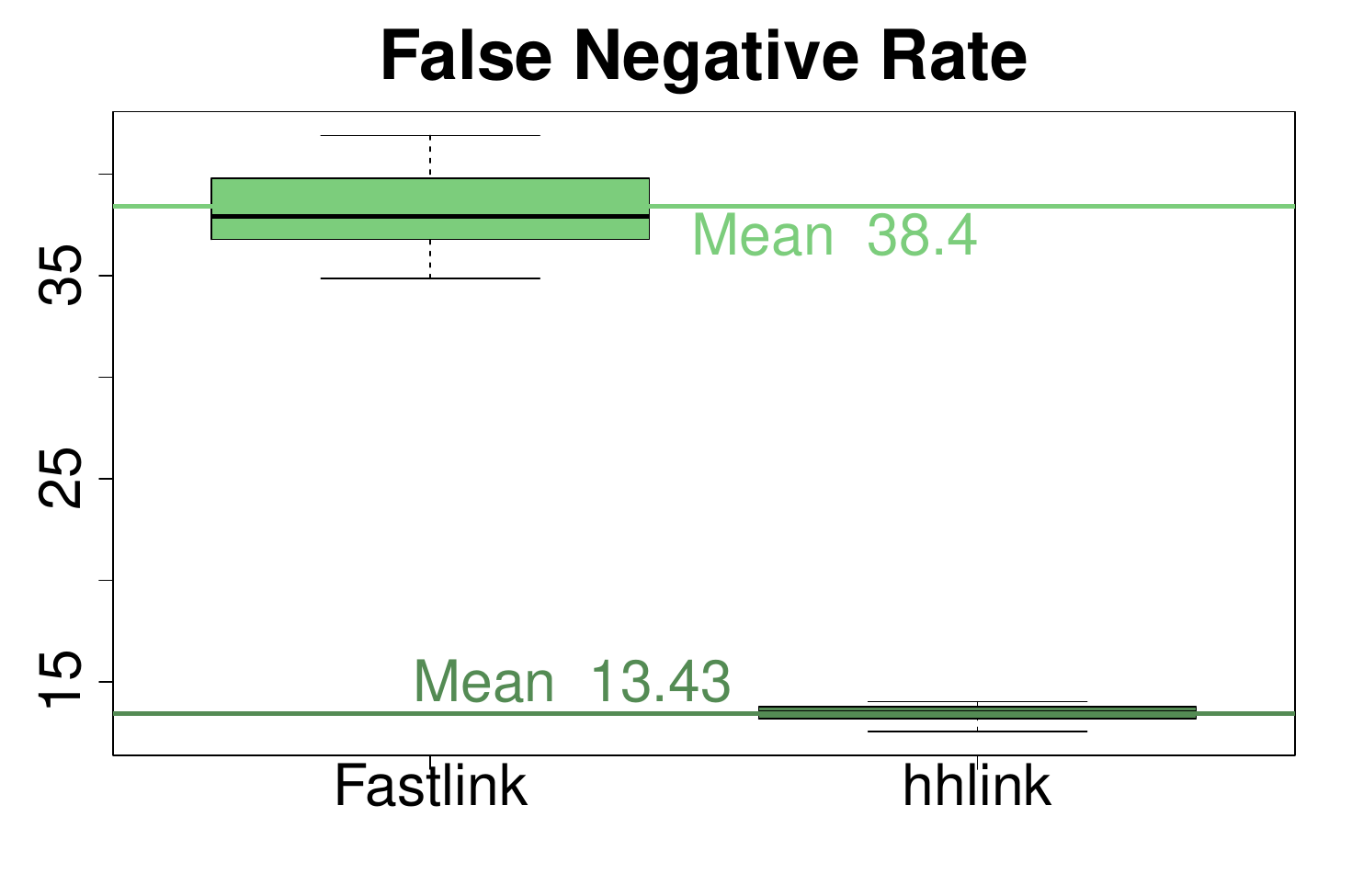}

  \centering
  \includegraphics[width=0.32\textwidth]{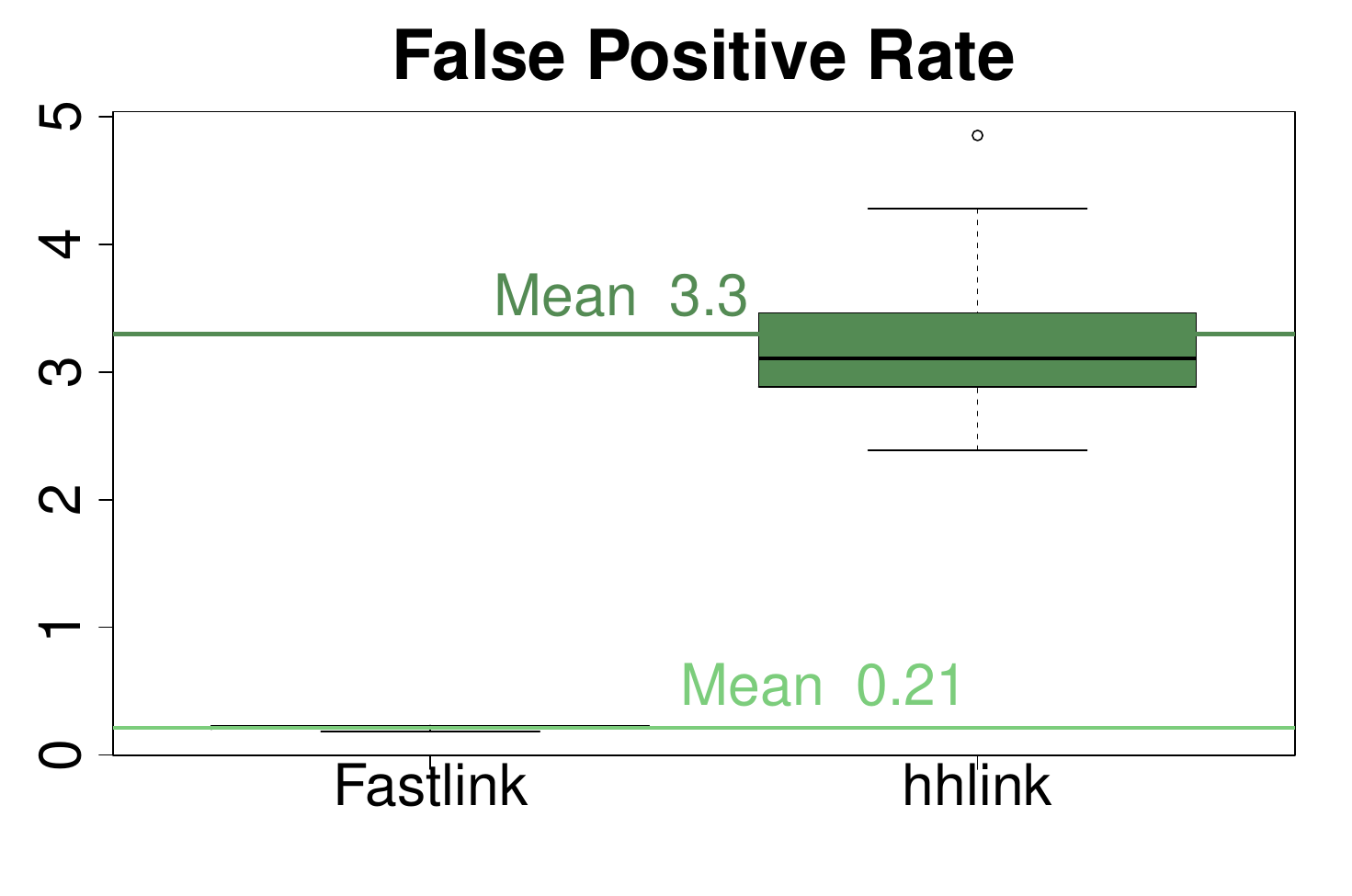}
  \centering
  \includegraphics[width=0.32\textwidth]{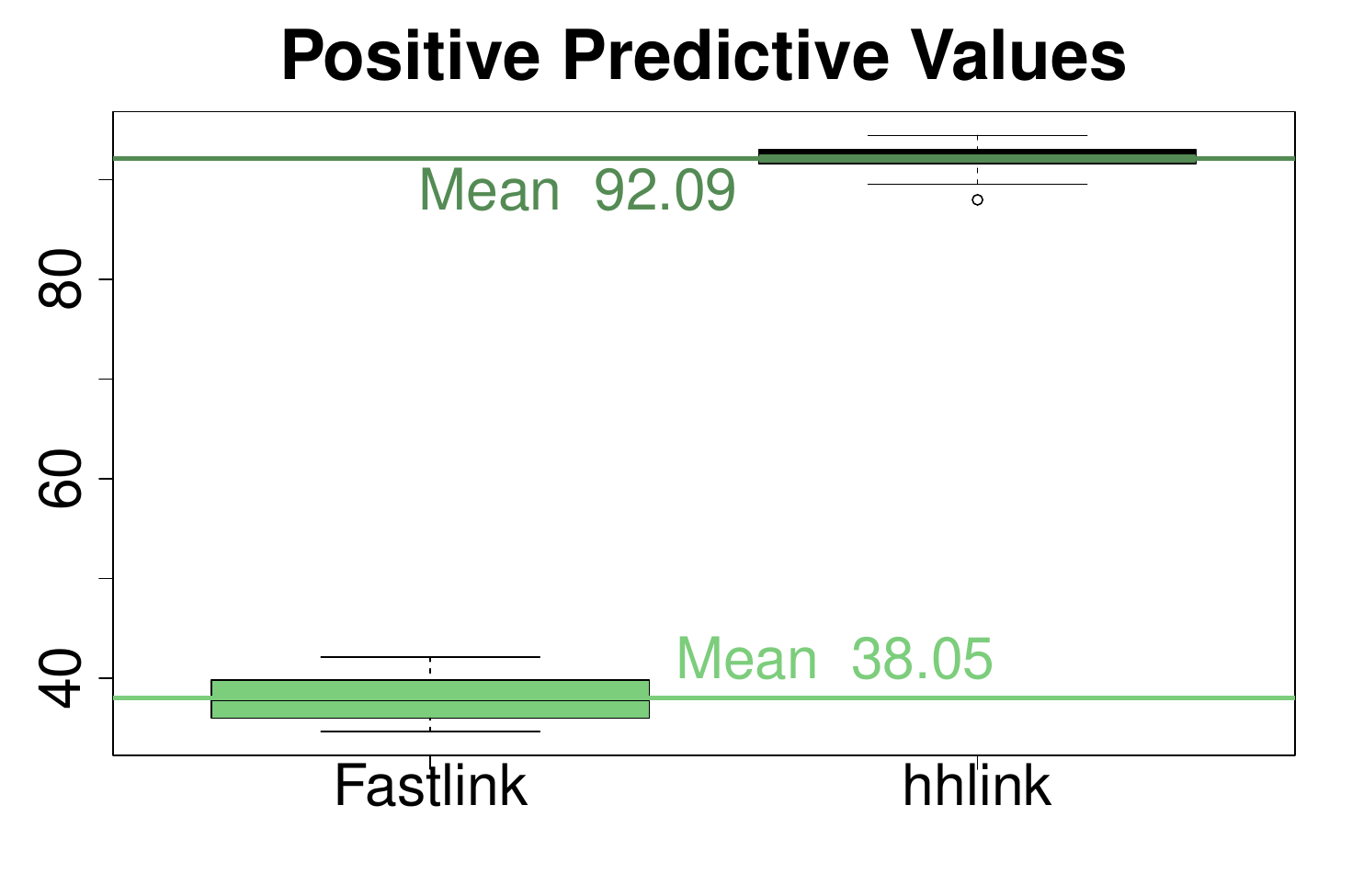}
  \centering
  \includegraphics[width=0.32\textwidth]{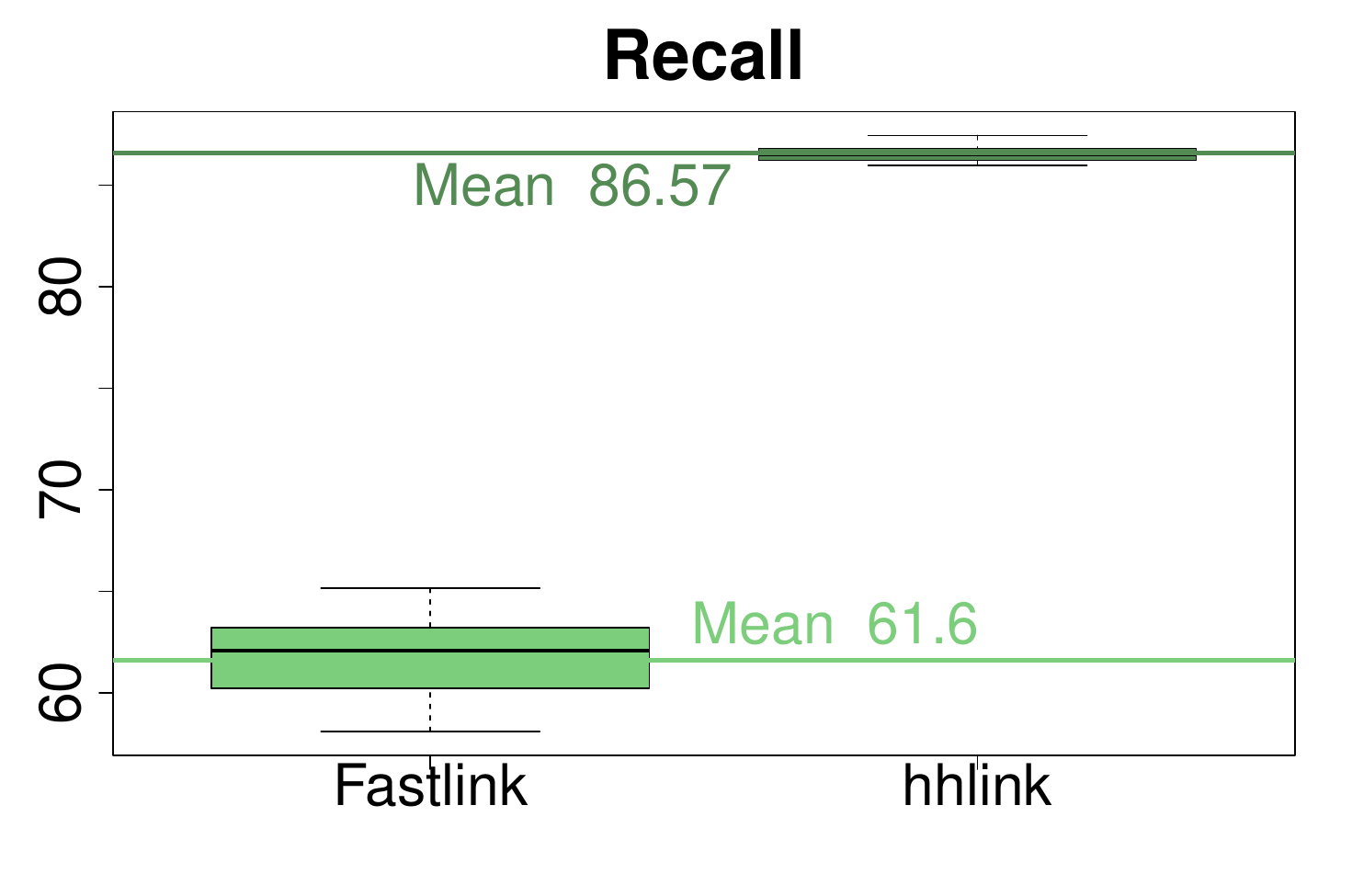}
\caption{Boxplots illustrating performance metrics for individual pair match status prediction on the test data for {\em \tt fastLink} and {\em \tt hhlink} methods.}
\label{f:samplesTest}
\end{figure}

As detailed in Section~\ref{s:comparison}, we compare our proposed {\tt hhlink}, which incorporates household information, with the \texttt{fastLink} method, which directly matches individuals. For each training and testing split, we compute performance metrics to assess the model's ability to correctly classify pairs as matches or non-matches. Figure~\ref{f:samplesTest} illustrates the boxplots of these performance measures for the test sets. 

As we examine the model's performance on the unseen test data, \texttt{hhlink} demonstrates superior performance across all metrics except for the false positive rate. The lower false positive rate exhibited by \texttt{fastLink} can be attributed to its propensity to assign only a limited number of pairs as matches. Consequently, it has a reduced chance of making false positive errors, as it predominantly categorizes matches as non-matches, incurring a higher false negative rate. The precision and recall values underscore the advantages of incorporating household information in the matching process. These values are significantly higher for \texttt{hhlink} compared to \texttt{fastLink}, which directly matches individuals. Specifically, the recall value indicates that \texttt{hhlink} correctly identifies approximately $87\%$ of pairs known to match, in contrast to the $62\%$ achieved by \texttt{fastLink}.

It is worth highlighting further that the number of pairs of individuals to be examined across databases is quite large, posing some computational burdens, since a large number of matching probabilities need to be estimated. Although blocking techniques can help alleviate this issue to some extent, \texttt{fastLink} still demands consideration of more than $3,000,000$ potential pairs on the training data and more than $1,500,000$ on the test data. In contrast, {\tt hhlink} offers an effective remedy by reducing the number of individual pairs requiring evaluation. Notably, the detection of matching households essentially serves as an additional blocking step. Consequently, one only needs to consider pairs between households that have been identified as matches, significantly reducing the number of pairs to be examined to around $33,000$ on the training set and $21,000$ on the test data.

Given the substantial volume of pairs for comparison, and the unbalanced nature of the framework in which most of the individual's pairs are non-matching pairs, good performance measures can also be achieved by a method that assigns most pairs as non-matching, even if the method is not well designed. Additionally, previous results are only accounting for pairs of individuals within matched households. Individuals inside households that were not matched have not been paired with any other individual, making it impossible to assess the model's ability to correctly classify individual pairs in those cases. Therefore, it is of interest to assess the performance of the record linkage methods at identifying if an individual in the $2014$ database has a match in the $2016$ database, regardless of the number of total pairs, and if the matched individual in $2016$ is correctly identified. In this case, we have two correct outcomes: the model correctly identifies the individual's match (Correct Matches) or it is able to correctly detect that the individual does not have a match (Correct Non-Matches).

\begin{figure}[b!]
  \centering
  \includegraphics[width=0.45\textwidth]{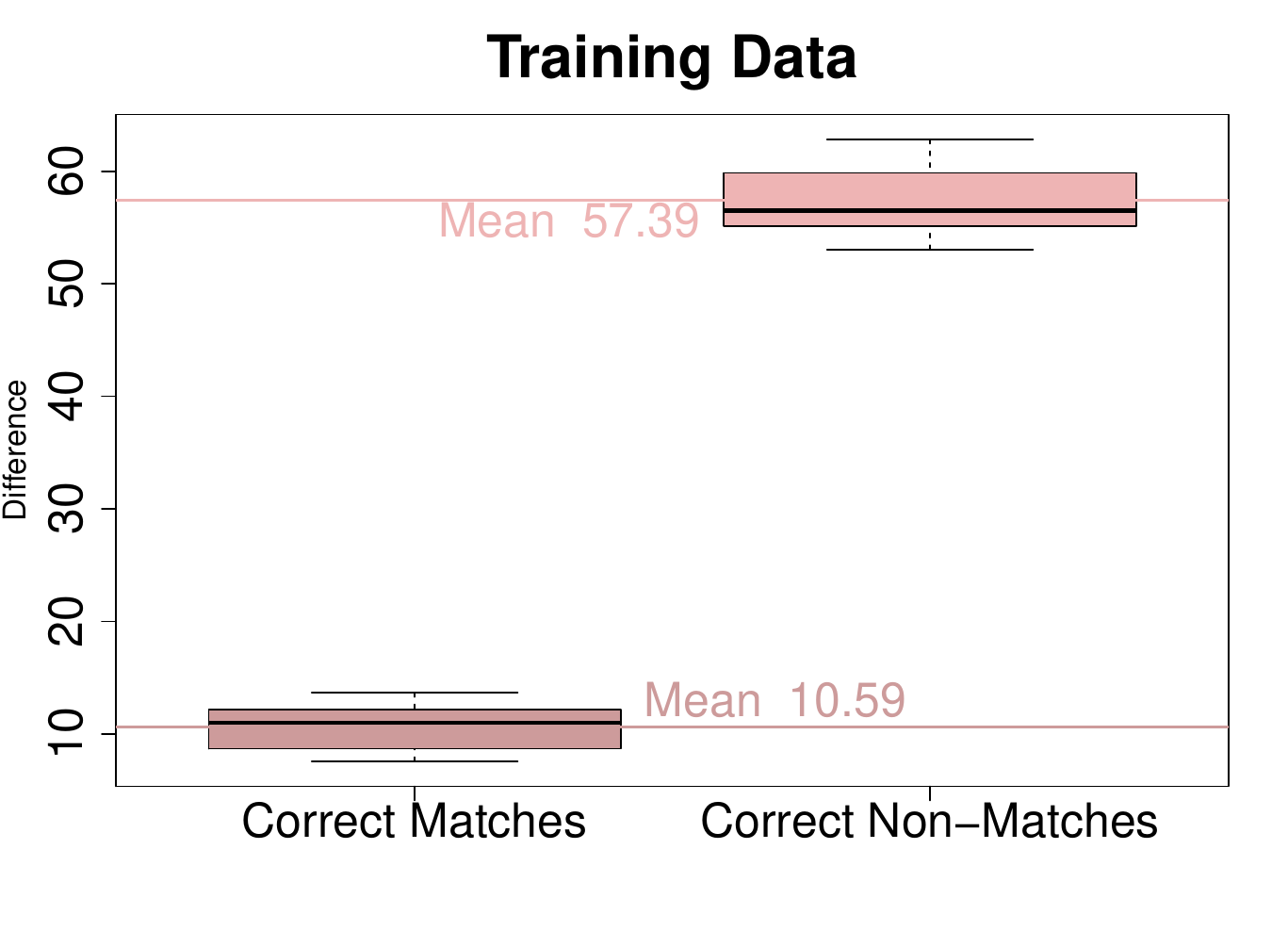}
  \centering
  \includegraphics[width=0.45\textwidth]{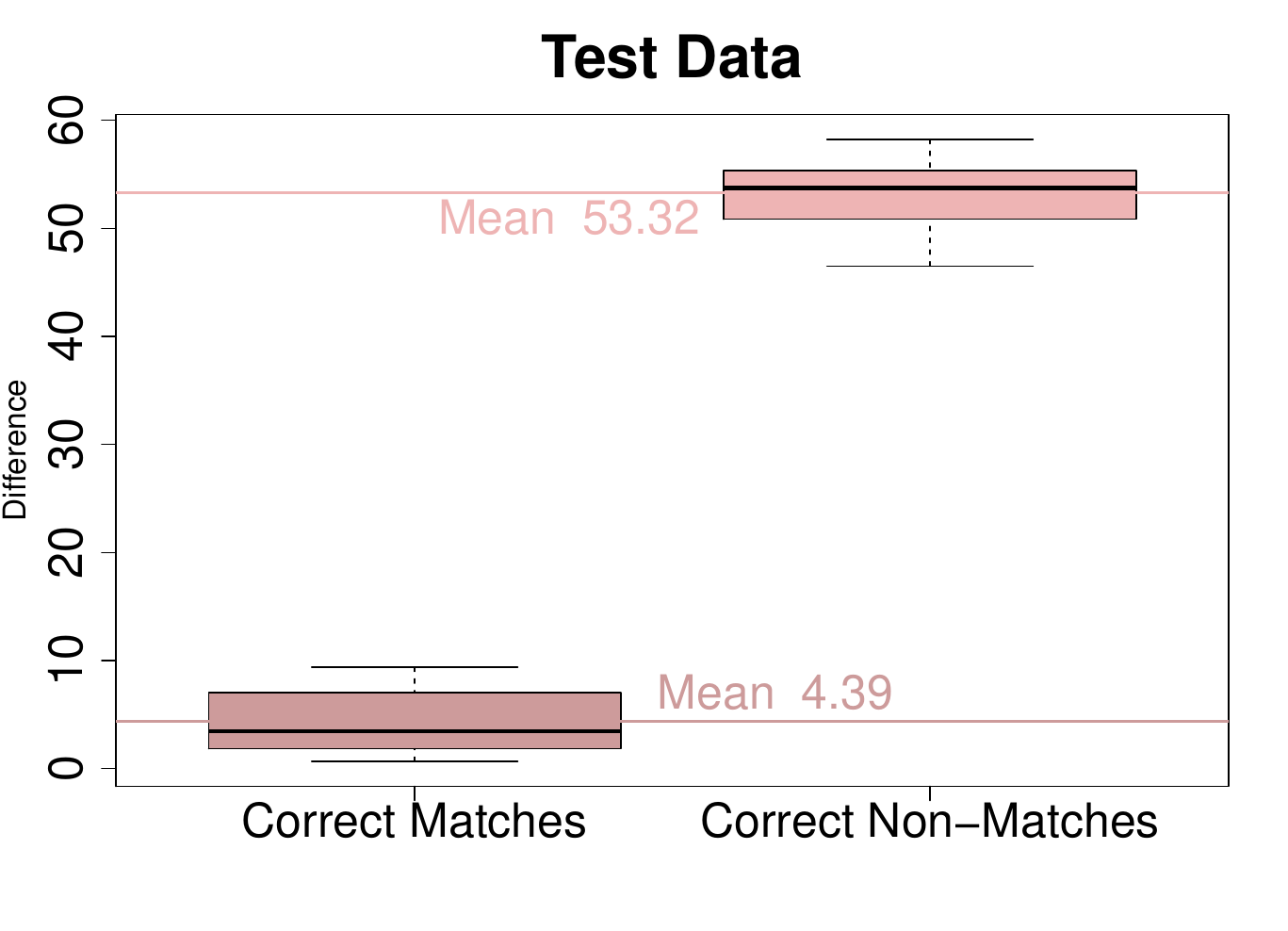}
\caption{Boxplot displaying the difference in percentages between correctly identified matches and correct non-matches between {\em \tt hhlink} and {\em \tt fastLink}. The plot represents the average point-to-point difference between these approaches, with positive values denoting the {\em \tt hhlink} better performance.}
\label{f:diff}
\end{figure}

In this regard, Figure~\ref{f:diff} presents boxplots illustrating the difference between the percentage of correct matches and correct non-matches identified by our \texttt{hhlink} approach in comparison to the \texttt{fastLink} method. In this figure, positive values indicate that {\tt hhlink}, incorporating household information, is better at identifying more correct matches or non-matches compared to {\tt fastLink}. Conversely, negative values suggest that \texttt{fastLink} performed better. An examination of the results reveals that, on average, \texttt{hhlink} exhibits a higher proportion of correctly identified matches in the training data, surpassing the \texttt{fastLink} proportion by $10.59$ percentage points. In the test data, this average difference is reduced to $4.39$ points. Notably, \texttt{hhlink} excels in accurately identifying individuals without matches, exhibiting a substantial average improvement of $57.39$ points in the training data and $53.32$ points in the test data when compared to the \texttt{fastLink} approach.

In summary, the results underscore the effectiveness of the \texttt{hhlink} method in correctly matching households. Additionally, the benefits of matching households become evident in the subsequent step of matching individuals, with consistently superior performance compared to methods directly matching individuals. 

These findings are derived from the split of the survey databases from $2014$ and $2016$ into various training and test sets. In the following section, we will present the results of a validation involving an external test data set, where the model is trained using complete surveys from $2014$ and $2016$, and then evaluated by matching the $2016$ survey with the $2020$ data.

\subsection{External Validation} \label{s:external}

This validation simulates a scenario where a new survey is available and a researcher is asked to match the individuals on this new database with the individuals on the previous survey. In this scenario, we will assess the method performance by using the surveys of $2014$ and $2016$ to train the method, then test it by matching the $2016$ survey with the new $2020$ data. 

\begin{table}[t]
\centering
\caption{Estimates of the household model parameters trained using the $2014$ and $2016$ survey data.}
\label{t:householdpar_external}
\begin{tabular}{lrrrrrrrrr}
    \hline \noalign{\smallskip}
                       & Intercept & SEX  & ANASC & CIT  & STUDIO & NACE & NASCREG & IREG & QUAL \\ \hline \noalign{\smallskip} 
    Estimate           & -0.69      & 2.86 & 14.76 & 0.00 & 1.60   & 1.42 & 3.35    & 7.15 & 0.00 \\ \noalign{\smallskip} \hline
\end{tabular}
\end{table}

Table~\ref{t:householdpar_external} provides the parameter estimates for the household model in this scenario. 
The estimates mirror those obtained in the internal validation, as detailed in Table~\ref{t:comparingParam}. The variable ANASC (year of birth) is the one with the highest estimated weight in the Hausdorff distance, while CIT (Italian citizenship) and QUAL (employment status) do not contribute to the distance. We note that in this instance the model is trained on the entirety of the $2014$ and $2016$ survey data, hence no standard deviation is associated with the estimates.

\begin{table}[b!]
\centering
\caption{Performance metrics for predicting the match status of household pairs. The {\em 2014-2016} column displays results from the training phase utilizing the entire $2014$ and $2016$ databases. The {\em 2016-2020} column reports results on the test scenario when matching the $2016$ database with the $2020$ data.}
\label{t:householdmodel_external}
\begin{tabular}{lrrrrr}
\hline\noalign{\smallskip}
Metric & 2014-2016 & 2016-2020 \\ 
\noalign{\smallskip}
\hline
\noalign{\smallskip}
$F_1$ Score        & 70.30 & 57.17 \\ 
FNR        & 27.66 & 45.56 \\ 
FPR      & 0.002  & 0.002 \\ 
PPV        & 68.37  & 60.19 \\ 
Recall & 72.34 & 54.44 \\ 
   \hline
\end{tabular}
\end{table}

Considering all possible pairs of households to be classified as matches or non-matches, Table~\ref{t:householdmodel_external} presents the performance measure for the household model considering the training (2014-2016) and test (2016-2020) scenario. The household model yields a positive predictive value (PPV) of $68.37\%$ when matching households from the $2014$ database with the $2016$ database, and a value of $60.19\%$ for matching households between the $2016$ and $2020$ surveys. These results indicate that, among all pairs classified as matches, the majority corresponds to true matches. Additionally, the high recall values suggest that the model is effective in identifying matching pairs. However, the results show an increase in the false negative rate in the test scenario. This discrepancy can be traced back to the threshold estimation process. The threshold is determined to achieve a proportion of estimated households in the training phase equal to the true percentage ($46.64\%$) of matching households between $2014$ and $2016$. However, when this threshold is applied to match households in the $2016$-$2020$ pair, which features a notably lower true proportion of matching households ($40.20\%$), the estimated threshold may be deemed too high for the more recent surveys. Consequently, this discrepancy contributes to an increase in the false negative rate.

\begin{table}[t!]
\centering
\caption{Ranking of correctly matching households by probability. The first column displays results from the training process using the $2014$ and $2016$ data, while the second column presents results for matching the $2016$ survey with the $2020$ data. A rank of $1$ signifies that the highest probability corresponds to the correct household match. Similarly, rank $2$ indicates the second-highest probability match, and so on.}
\label{t:positionmodel_external}
\begin{tabular}{rrrrrr}
\hline\noalign{\smallskip}
Rank & 2014-2016 & 2016-2020 \\ 
\noalign{\smallskip}
\hline
\noalign{\smallskip}
1        & 74.21 & 64.06 \\ 
2        & 5.16 & 5.49 \\ 
3        & 1.97  & 2.43 \\ 
4        & 1.39  & 2.18 \\ 
$\geq$ 5 & 17.26 & 25.85 \\ 
   \hline
\end{tabular}
\end{table}

Table~\ref{t:positionmodel_external} provides the rank analysis for the household model. The rank serves as a metric indicating the position of the true match within the household matching process.
The results consistently echo the previous findings obtained in the internal validation. In both scenarios, the highest match probability is predominantly associated with the true match in the subsequent survey for the majority of households. 
In particular, when matching households from $2016$ to $2020$, in $64.06\%$ of cases the highest probability is associated with the true match. In general, across both scenarios, the true household match tends to fall within the top three highest match probabilities. However, for the $2014-2016$ case, there are instances ($17.36\%$) where a household's probability of a match with its true match exceeds rank $4$, while for the $2016-2020$ case, this occurs in $25.85\%$ of cases. As the scenario of training the model on the 2014-2016 databases and matching the 2016 survey to the 2020 data is more challenging, the likelihood of including the actually matching household in the top four positions is reduced.

\begin{table}[b]
\centering
\caption{Estimates of the individual model parameters trained using the data of $2014$ and $2016$}
\label{t:indpar_external}
\begin{tabular}{lrrrrrrrrr}
    \hline \noalign{\smallskip}
                       & Intercept & SEX  & ANASC & CIT  & STUDIO & NACE & NASCREG & IREG & QUAL \\ \hline \noalign{\smallskip} 
    Estimate           & 1.88      & 2.77 & 4.43 & 0.50 & 1.17   & 0.47 & 0.95    & 0.00 & 0.58 \\ \noalign{\smallskip} \hline
    \end{tabular}
\end{table}

After the household matching step across databases, the subsequent step involves fitting the individual model given the estimated household match status. Table~\ref{t:indpar_external} provides the parameter estimates for the individual model, which is trained using pairs of individuals within the matched households from the $2014$ and $2016$ surveys.
The estimates are in line with those reported in the internal validation, where the model was trained only on random subsets of the $2014$ and $2016$ databases, as reported in Table~\ref{t:individualparam}. Also in this case ANASC and SEX have the largest weights, and IREG does not exert a significant impact in matching individuals. This consistency underscores the robustness of the estimates across different evaluation scenarios. 

As elaborated in Section~\ref{s:linearprog}, following the estimation of the parameters in the individual model, we compute the probability of a match for all the pairs of individuals. Subsequently, leveraging a linear programming framework, we link individuals across databases. We apply {\tt hhlink} to match pairs of individuals across the $2014$ and $2016$ surveys in the training phase, as well as link individuals between the $2016$ and $2020$ surveys in the testing phase. We also use the \texttt{fastLink} approach to match individuals across the databases for comparison. We remark that, in applying \texttt{fastLink}, blocking is employed to mitigate computational workload, as detailed in Section~\ref{s:comparison}. The match performance in each of these scenarios is reported in Table~\ref{t:performance}.

\begin{table}[t!]
\centering
\caption{Comparison of individuals matching quality between {\em \texttt{hhlink}} and {\em \texttt{fastLink}}.}
\label{t:performance}
\begin{tabular}{lrrrr}
\hline\noalign{\smallskip}
  & \multicolumn{2}{c}{\texttt{hhlink}} & \multicolumn{2}{c}{\texttt{fastLink}}\\
  & 2014-2016 & 2016-2020 & 2014-2016 & 2016-2020\\
         \hline\noalign{\smallskip}
$F_1$ Score & 87.29 & 82.93 & 30.37 & 26.57  \\
\noalign{\smallskip} 
FNR & 13.25 & 16.69 & 57.47 & 61.88   \\
\noalign{\smallskip} 
FPR      & 5.05 & 7.44 & 0.12 & 0.12 \\
\noalign{\smallskip} 
PPV      & 87.84 & 82.55  & 23.62  & 20.03  \\
\noalign{\smallskip} 
Recall       & 86.75 &  83.31  & 42.53 & 38.12   \\ 
\hline\noalign{\smallskip}
\end{tabular}
\end{table}

The \texttt{hhlink} approach has an $F_1$ score higher than \texttt{fastLink} in both linking the 2014 with the 2016 survey and the $2016$ and $2020$ database. This indicates that the record linkage of individuals based on matched household information outplays \texttt{fastLink} which matches individuals directly. As before, the lower false positive rate (FPR) of {\tt fastLink} is associated with fewer matches being assigned in this method, resulting in a higher false negative rate (FNR). Finally, the precision and recall values reinforce that the inclusion of household information is beneficial to the process of matching individuals since such values for {\tt hhlink} are more than double than those of \texttt{fastLink}.

As before, the extensive number of individual pairs across multiple databases presents significant computational challenges, especially when dealing with complete sets of databases from $2014$, $2016$, and $2020$, reaching evaluation of $7,000,000$ pairs in the \texttt{fastLink}. Such a large volume of pairs can potentially skew performance assessments. Therefore, it's crucial to assess record linkage methods based on their ability to accurately identify matches in the $2016$ database for individuals in $2014$, or matches in the $2020$ database for individuals in $2016$, regardless of the total number of pairs. This evaluation focuses on two outcomes: correctly identifying matches (Correct Matches) and accurately discerning non-matches (Correct Non-Matches). Table~\ref{t:correctmatches} provides a comprehensive overview of these performance metrics for the examined record linkage approaches.

\begin{table}[b]
\centering
\caption{Number of correctly detected individual matches and non-matches for {\em \tt hhlink} and {\em \tt fastLink}.}
\label{t:correctmatches}
\begin{tabular}{lrrrr}
\hline\noalign{\smallskip}
  & \multicolumn{2}{c}{\texttt{hhlink}} & \multicolumn{2}{c}{\texttt{fastLink}}\\
  & 2014-2016 & 2016-2020 & 2014-2016 & 2016-2020\\
                    \hline\noalign{\smallskip}
Correct Matches     & 5578 (64.41\%) & 2985 (46.39\%) & 3659 (42.25\%) & 2420 (37.61\%) \\
\noalign{\smallskip} 
Correct Non-Matches & 10125 (94.57\%) & 9615 (95.88\%) & 3103 (28.98\%) & 3473 (34.63\%)\\  
\hline\noalign{\smallskip}
\end{tabular}
\end{table}

The \texttt{hhlink} approach consistently outperforms the direct individual linkage approach. When household information is factored in, it results in the detection of $5578$ (or $64.41\%$) of the $8660$ actual matches between individuals in the $2014$ and $2016$ datasets and $2985$ ($46.39\%$) of $6434$ for the individuals in the $2016$ and $2020$ datasets. In stark contrast, the \texttt{fastLink} approach lags behind with detection of only $3659$ ($42.25\%$) and $2420$ ($37.61\%$) respectively. It's crucial to note that the \texttt{hhlink} approach can correctly identify more than $90\%$ of the individuals that do not have a match while the percentage is around $30\%$ for the \texttt{fastLink} approach. We highlight that the reduced performance when matching the $2016$ survey with the $2020$ database in the testing phase, in comparison to the internal validation results, may be attributed to the amends done in $2020$ in the traditional sampling design to improve the sample representativeness of some population groups, as informed in the Bank of Italy website. The four-year gap between these surveys, instead of the usual two years can also contribute to this. Nonetheless, these findings underscore the significant enhancement achieved by incorporating household information, leading to an increased number and quality of correctly identified matches.

\section{Discussion and Conclusions}
\label{s:conclusion}

This work introduced a novel record linkage approach, {\tt hhlink}, contributing to two key aspects. Firstly, it introduces the Hausdorff distance as a valuable metric for effectively measuring the dissimilarity between households during the matching process. Secondly, it underscores the advantages of initiating the matching process at the household level when linking individual records across databases, ultimately improving data integration and the quality of the matched results.

The proposed {\tt hhlink} approach is a multi-step methodology. The first step employs the Hausdorff distance to estimate the probability of a match between pairs of households, based on linear combinations of distances between individual features. The following step employs logistic regression and linear programming optimization to match individual records within identified matched households. 

The {\tt hhlink} method is showcased and evaluated in application to record linkage of the Italian Survey of Household Income and Wealth (SHIW) data, demonstrating the substantial benefits of considering household information when linking individual records across databases. 
Across internal and external validation frameworks, evaluation metrics consistently indicate superior performance of {\tt hhlink} compared to a method that directly matches individual records without leveraging household information. 

A limitation of the proposed approach is in the supervised nature of the method, which requires the availability of labeled data where identifiers of matching households and individuals between databases are needed for training.
This opens interesting avenues for future research. Future work will explore extensions to unsupervised learning for record linkage, where grouping information of the instances is available but not identifiers that can be used for matching. Unsupervised extensions of the proposed approach could be particularly useful in matching surveys with a larger time gap.

The proposed approach has been developed in application to record linkage of survey data collected at the household level. However, we remark that the proposed framework holds the potential for record linkage in other databases with grouping and hierarchical structures. The methodology's applicability extends beyond the specific data used, making it a valuable tool for data integration and analysis in various domains where grouping information on the individual records to be linked is available.
 
\section*{Declarations}
\textbf{Funding:} This publication has emanated from research conducted with the financial support of Science Foundation Ireland under grant numbers 18/CRT/6049 and 12/RC/2289\_P2 and a visiting period at Collegium de Lyon.

\noindent \textbf{Conflicts of interest:} The authors declare that there is no conflict of interest.

\noindent \textbf{Ethical Conduct:} The manuscript is only submitted to the Journal of Classification. The submitted work is
original and is not published elsewhere in any form or language.

\noindent \textbf{Data Availability:} The data that support the findings of this study are openly available on the Bank of Italy website \citep{URL}.

\newpage

\bibliographystyle{apalike}
\bibliography{sample.bib}

\begin{thebibliography}{}

\bibitem[Abramitzky et~al., 2021]{abramitzky:2021}
Abramitzky, R., Boustan, L., Eriksson, K., Feigenbaum, J., and Pérez, S.
  (2021).
\newblock Automated linking of historical data.
\newblock {\em Journal of Economic Literature}, 59(3):865--918.

\bibitem[Abramitzky et~al., 2020]{abramitzky:2020}
Abramitzky, R., Mill, R., and Pérez, S. (2020).
\newblock Linking individuals across historical sources: A fully automated
  approach.
\newblock {\em Historical Methods: A Journal of Quantitative and
  Interdisciplinary History}, 53(2):94--111.

\bibitem[Albert and Anderson, 1984]{albert1984existence}
Albert, A. and Anderson, J.~A. (1984).
\newblock On the existence of maximum likelihood estimates in logistic
  regression models.
\newblock {\em Biometrika}, 71(1):1--10.

\bibitem[{Bank of Italy}, 2022]{URL}
{Bank of Italy} (2022).
\newblock Bilanci delle famiglie {I}taliane.
\newblock
  \url{https://www.bancaditalia.it/statistiche/tematiche/indagini-famiglie-imprese/bilanci-famiglie/documentazione/index.html}
  (Accessed: 2022-10-11 and 2023-08-03).

\bibitem[Biancotti et~al., 2008]{MesurementErrorSHIW}
Biancotti, C., D'Alessio, G., and Neri, A. (2008).
\newblock Measurement error in the {B}ank of {I}taly's {S}urvey of {H}ousehold
  {I}ncome and {W}ealth.
\newblock {\em Review of Income and Wealth}, 54(3):466--493.

\bibitem[Cohen et~al., 2003]{cohen2003comparison}
Cohen, W.~W., Ravikumar, P., and Fienberg, S.~E. (2003).
\newblock A comparison of string distance metrics for name-matching tasks.
\newblock In {\em Proceedings of the 2003 International Conference on
  Information Integration on the Web}, pages 73–--78. AAAI Press.

\bibitem[Eiter and Mannila, 1997]{eiter1997distance}
Eiter, T. and Mannila, H. (1997).
\newblock Distance measures for point sets and their computation.
\newblock {\em Acta Informatica}, 34(2):109--133.

\bibitem[Enamorado et~al., 2019]{enamorado2019using}
Enamorado, T., Fifield, B., and Imai, K. (2019).
\newblock Using a probabilistic model to assist merging of large-scale
  administrative records.
\newblock {\em American Political Science Review}, 113(2):353–371.

\bibitem[Enamorado et~al., 2020]{fastLink}
Enamorado, T., Fifield, B., and Imai, K. (2020).
\newblock {\em fastLink: Fast Probabilistic Record Linkage with Missing Data}.
\newblock R package version 0.6.0.

\bibitem[Fellegi and Sunter, 1969]{fellegi1969theory}
Fellegi, I.~P. and Sunter, A.~B. (1969).
\newblock A theory for record linkage.
\newblock {\em Journal of the American Statistical Association},
  64(328):1183--1210.

\bibitem[Fortini et~al., 2001]{fortini2001bayesian}
Fortini, M., Liseo, B., Nuccitelli, A., and Scanu, M. (2001).
\newblock On {B}ayesian record linkage.
\newblock {\em Research in Official Statistics}, 4(1):185--198.

\bibitem[Friedman et~al., 2010]{glmnet_paper}
Friedman, J., Hastie, T., and Tibshirani, R. (2010).
\newblock Regularization paths for generalized linear models via coordinate
  descent.
\newblock {\em Journal of Statistical Software}, 33(1):1--22.

\bibitem[Friedman et~al., 2021]{glmnet}
Friedman, J., Hastie, T., Tibshirani, R., Narasimhan, B., Tay, K., Simon, N.,
  Qian, J., and Yang, J. (2021).
\newblock {\em glmnet: Lasso and Elastic-Net Regularized Generalized Linear
  Models}.
\newblock R package version 4.1-1.

\bibitem[Frisoli and Nugent, 2018]{frisoli2018exploring}
Frisoli, K. and Nugent, R. (2018).
\newblock Exploring the effect of household structure in historical record
  linkage of early 1900s {I}reland census records.
\newblock In {\em 2018 IEEE International Conference on Data Mining Workshops
  (ICDMW)}, pages 502--509.

\bibitem[Fu et~al., 2011]{fu2011automatic}
Fu, Z., Christen, P., and Boot, M. (2011).
\newblock Automatic cleaning and linking of historical census data using
  household information.
\newblock In {\em 2011 IEEE 11th International Conference on Data Mining
  Workshops}, pages 413--420.

\bibitem[Fu et~al., 2014]{fu2014graph}
Fu, Z., Christen, P., and Zhou, J. (2014).
\newblock A graph matching method for historical census household linkage.
\newblock In Tseng, V.~S., Ho, T.~B., Zhou, Z.-H., Chen, A. L.~P., and Kao,
  H.-Y., editors, {\em Advances in Knowledge Discovery and Data Mining}, pages
  485--496. Springer International Publishing.

\bibitem[Hausdorff, 1914]{hausdorffbook}
Hausdorff, F. (1914).
\newblock {\em Grundz{\"u}ge der Mengenlehre}.
\newblock Leipzig, Von Veit.

\bibitem[Heinze, 2006]{heinze2006comparative}
Heinze, G. (2006).
\newblock A comparative investigation of methods for logistic regression with
  separated or nearly separated data.
\newblock {\em Statistics in Medicine}, 25(24):4216--4226.

\bibitem[Helgertz et~al., 2022]{helgertz:2022}
Helgertz, J., Price, J., Wellington, J., Thompson, K.~J., Ruggles, S., and
  Fitch, C.~A. (2022).
\newblock A new strategy for linking {U.S.} historical censuses: A case study
  for the {IPUMS} multigenerational longitudinal panel.
\newblock {\em Historical Methods: A Journal of Quantitative and
  Interdisciplinary History}, 55(1):12--29.

\bibitem[Herzog et~al., 2007]{herzog2007data}
Herzog, T.~N., Scheuren, F.~J., and Winkler, W.~E. (2007).
\newblock {\em Data Quality and Record Linkage Techniques}, volume~1.
\newblock Springer.

\bibitem[Moretti et~al., 2019]{moretti2019optimization}
Moretti, D., Valentino, L., and Tuoto, T. (2019).
\newblock Optimization routines for enforcing one-to-one matches in record
  linkage problems.
\newblock {\em The R Journal}, 11(1):185.

\bibitem[Nash, 2014]{optimx2}
Nash, J.~C. (2014).
\newblock On best practice optimization methods in {R}.
\newblock {\em Journal of Statistical Software}, 60(2):1–14.

\bibitem[Nash and Varadhan, 2011]{optimx1}
Nash, J.~C. and Varadhan, R. (2011).
\newblock Unifying optimization algorithms to aid software system users: optimx
  for {R}.
\newblock {\em Journal of Statistical Software}, 43(9):1–14.

\bibitem[Nash et~al., 2022]{optimx}
Nash, J.~C., Varadhan, R., and Grothendieck, G. (2022).
\newblock {\em optimx: Expanded Replacement and Extension of the `optim'
  Function}.
\newblock R package version 2022-4.30.

\bibitem[On et~al., 2007]{on2007group}
On, B.-W., Koudas, N., Lee, D., and Srivastava, D. (2007).
\newblock Group linkage.
\newblock In {\em 2007 IEEE 23rd International Conference on Data Engineering},
  pages 496--505.

\bibitem[Papadakis et~al., 2022]{papadakis2022bipartite}
Papadakis, G., Efthymiou, V., Thanos, E., and Hassanzadeh, O. (2022).
\newblock Bipartite graph matching algorithms for clean-clean entity
  resolution: An empirical evaluation.
\newblock In {\em Proceedings of the 25th International Conference on Extending
  Database Technology (EDBT)}, pages 462--474.

\bibitem[{R Core Team}, 2022]{rsoftware}
{R Core Team} (2022).
\newblock {\em R: A Language and Environment for Statistical Computing}.
\newblock R Foundation for Statistical Computing, Vienna, Austria.

\bibitem[Ruggles et~al., 2018]{ruggles:2018}
Ruggles, S., Fitch, C.~A., and Roberts, E. (2018).
\newblock Historical census record linkage.
\newblock {\em Annual Review of Sociology}, 44(1):19--37.

\bibitem[Sadinle, 2017]{sadinle2017bayesian}
Sadinle, M. (2017).
\newblock Bayesian estimation of bipartite matchings for record linkage.
\newblock {\em Journal of the American Statistical Association},
  112(518):600--612.

\bibitem[Sadinle and Fienberg, 2013]{sadinle2013generalized}
Sadinle, M. and Fienberg, S.~E. (2013).
\newblock A generalized {F}ellegi–{S}unter framework for multiple record
  linkage with application to homicide record systems.
\newblock {\em Journal of the American Statistical Association},
  108(502):385--397.

\bibitem[Sayers et~al., 2015]{sayers2016probabilistic}
Sayers, A., Ben-Shlomo, Y., Blom, A.~W., and Steele, F. (2015).
\newblock Probabilistic record linkage.
\newblock {\em International Journal of Epidemiology}, 45(3):954--964.

\bibitem[Steorts et~al., 2016]{steorts2016bayesian}
Steorts, R.~C., Hall, R., and Fienberg, S.~E. (2016).
\newblock A {B}ayesian approach to graphical record linkage and deduplication.
\newblock {\em Journal of the American Statistical Association},
  111(516):1660--1672.

\bibitem[Steorts et~al., 2014]{comparisonblocking}
Steorts, R.~C., Ventura, S.~L., Sadinle, M., and Fienberg, S.~E. (2014).
\newblock A comparison of blocking methods for record linkage.
\newblock In Domingo-Ferrer, J., editor, {\em Privacy in Statistical
  Databases}, pages 253--268. Springer International Publishing.

\bibitem[Tancredi and Liseo, 2011]{tancredi:liseo:2011}
Tancredi, A. and Liseo, B. (2011).
\newblock {A hierarchical Bayesian approach to record linkage and population
  size problems}.
\newblock {\em The Annals of Applied Statistics}, 5(2B):1553--1585.

\bibitem[Winkler, 1990]{winkler1990string}
Winkler, W.~E. (1990).
\newblock String comparator metrics and enhanced decision rules in the
  {F}ellegi-{S}unter model of record linkage.
\newblock In {\em Proceedings of the Section on Survey Research Methods}, pages
  354--359. American Statistical Association.

\end{thebibliography}

\end{document}